\documentclass[14pt,preprint]{aastex}
\DeclareMathAlphabet\mathbfcal{OMS}{cmsy}{b}{n}

\usepackage{amsmath}
\usepackage{amssymb}
\usepackage{graphicx,graphics}
\usepackage{placeins}
\usepackage[english]{babel}
\usepackage{natbib}
\usepackage{color}
\usepackage{subfigure}
\usepackage{multirow}
\usepackage{enumerate}

\newcommand{\be}{\begin{equation}}
\newcommand{\ee}{\end{equation}}
\newcommand{\out}{\rm per}
\newcommand{\aout}{a_{\out}}
\newcommand{\bout}{b_{\out}}
\newcommand{\eout}{e_{\out}}
\newcommand{\fout}{f_{}}
\newcommand{\Pout}{P_{\rm out}}
\newcommand{\brout}{\bfr_{\out}}
\newcommand{\hrout}{\mathbf{\hat r}_{\out}}
\newcommand{\rout}{r_{\out}}

\newcommand{\ain}{a}
\newcommand{\mper}{m_{\rm per}}
\newcommand{\tsec}{t_{\rm sec}}

\newcommand{\Phiper}{\Phi_{\rm per}}
\newcommand{\PhiQ}{\Phi_{\rm Quad}}
\newcommand{\PhiO}{\Phi_{\rm Oct}}
\def\bfj{\,\textbf{j}}
\def\sbfj{\textbf{j}}
\def\bfe{\textbf{e}}

\def\bfr{{\bf r}}
\def\epoct{{\epsilon_{Oct}}}
\def\epCDA{{\epsilon_{SA}}}
\def\cD{\mathcal{D}}

\begin{document}

\title{Double-Averaging Can Fail to Characterize the Long-Term Evolution of Lidov-Kozai Cycles \\ and
\\ Derivation of an Analytical Correction}

\author{Liantong Luo\altaffilmark{1},
        Boaz Katz\altaffilmark{2}, Subo Dong\altaffilmark{1}}

\altaffiltext{1}{Kavli Institute for Astronomy and Astrophysics, Peking University, Yi He Yuan Road 5, Hai Dian District, Beijing 100871, China}
\altaffiltext{2}{Department of Particle Physics and Astrophysics, Weizmann Institute of Science, Rehovot 76100, Israel}
\shorttitle{ Lidov-Kozai Double-Averaging Correction}
\shortauthors{Luo & Katz & Dong}

\begin{abstract}
The double-averaging (DA) approximation is widely employed 
as the standard technique in studying the secular evolution of the 
hierarchical three-body system. We show that 
effects stemmed from the short-timescale oscillations ignored 
by DA can accumulate over long timescales and lead to significant errors 
in the long-term evolution of the Lidov-Kozai cycles.
 In particular, the conditions for having an orbital flip, where the inner orbit
 switches between prograde and retrograde with respect to the outer orbit and 
 the associated extremely high eccentricities during the switch, can be modified significantly. The failure of DA can arise for a relatively strong perturber where the mass of the tertiary is considerable compared to the total mass of the inner binary. This issue can be relevant for astrophysical systems such as stellar triples, planets in stellar binaries, stellar-mass binaries orbiting massive black holes and moons of the planets perturbed by the Sun. We derive analytical equations for the short-term oscillations of the inner orbit to the leading order for all inclinations, eccentricities and mass ratios. Under the test particle approximation, we derive the ``corrected double-averaging'' (CDA) equations by incorporating the effects of short-term oscillations into the DA. By comparing to N-body integrations, we show that the CDA equations successfully correct most of the errors of the long-term evolution under the DA approximation for a large range of initial conditions. We provide an implementation of CDA that can be directly added to codes employing DA equations.
\end{abstract}

\keywords{gravitation -- methods: analytical and numerical}

\section{Introduction}\label{sec:Introduction}
The long-standing three-body problem was initially motivated by
studying planetary motions in the Solar System. The discoveries
of extrasolar planet systems with rich orbital architectures have
recently reinvigorated the research into this problem
\citep[see, e.g.,][]{innanen97, mazeh97, holman97}. Unlike the nearly 
circular and coplanar planetary orbits in the Solar System, 
some exoplanets are found on eccentric
and/or inclined orbits (see review by \citealt{winnaraa} and references
therein).
In recent years, dynamical processes involving highly eccentric orbits
have been invoked to interpret a wide array of astrophysical
phenomena \citep[see, e.g.][]{mazeh79, kiseleva98, wumurray, warmjupiter,
kozaibh, kozaibh2, bodewegg, kozaiasteroid, kozaibluestraggler, thompson11, katzdong}. Highly eccentric orbits can bring
two of the bodies to close approaches near the pericenters that result in
dissipative interactions, mergers or collisions. In particular, a popular class of mechanisms to
explain the formation of observed short-period (order of $\sim$day) planetary and stellar orbits invokes tidal dissipation during close encounters in three-body systems \citep[see, e.g.][]{mazeh79, kiseleva98, fabryckytremeine, dong13, wumurray, ford00, naoz11, katz11, lithwicknaoz11, socrates12, dawson15}. It has also been recently realized that the rate for mergers and collisions of white-dwarfs (WDs) can be significantly enhanced in field triple systems \citep{thompson11, katzdong} and
quadrupole systems \citep{pejcha}, and WD collisions in triples may be responsible for the majority of type Ia supernovae \citep{katzdong, kushnir13, dong15}.

The orbital evolution of a particular three-body system for a given set of
initial parameters can be accurately calculated using a computer with direct
integration of the inverse-square law. While such calculations play a crucial role in studying the three-body problem, analytic approximations have turned out to be equally important by allowing a deeper understanding as well as providing efficient routes for calculating the evolution for a large ensemble of initial parameters. In fact, for many cases in relevant astrophysical settings, direct numerical integrations are prohibitive due to the large amount of initial conditions needed to be scanned and large numbers of orbits (thousands, millions or even billions) to compute.

Many powerful analytical tools have been developed over
centuries to study the nearly circular and coplanar Solar System orbits,
but many of them are not applicable for highly-eccentric
and inclined orbits.
A pathbreaking analytical insight was achieved about half
a century ago by \citet{lidov62} and \citet{kozai62}, who solved the 3-body problem analytically at all eccentricities and inclinations
in the limit of high hierarchy - an inner binary, orbited by a distant third body (the perturber). They found that an initially nearly circular orbit of
the inner binary can be excited to high eccentricity by an inclined perturber over long timescales (i.e., secular timescales).
A large class of astrophysical-relevant systems are hierarchical for the simple reason that if they are not, one of the bodies can be ejected on a short time scale. The Lidov-Kozai solution stands as a starting point for a wide range of studies involving high eccentricities and inclinations.

In the limit of high hierarchy, on short timescales (timescales comparable to the
orbital periods), the time evolution of the system
can be well described by two separate Keplerian orbits -- 1) the (inner) orbit of the inner binary; 2) the outer orbit of the perturber orbiting the inner binary's center of mass. Over long timescales (longer than the outer orbital period), the two orbits exchange angular momentum periodically and their eccentricities and mutual inclinations oscillate, which are called the Lidov-Kozai cycles. The exchange of energy between the orbits is negligible over long timescale, and thus their semi-major axis values $\ain$ and $\aout$ are practically fixed in time. The evolution can be calculated analytically by expanding the interaction Hamiltonian to the leading (quadrupole) term in the small parameter $\ain/\aout$ and averaging the equations of motion over the orbits. Averaging over the inner orbit only is called ``single-averaging'' (SA) and over both the inner and outer orbit is called ``double-averaging'' (DA). \citet{lidov62} and \citet{kozai62} arrived at their solution by 
doing DA, and following their works, DA has since been widely used   
as the standard method to study and apply the Lidov-Kozai solution.

Throughout this work, we focus on the simplifying case that one component of the inner binary has negligible mass (the test particle limit). Our results are also applicable to a binary system with comparable masses orbiting a much more massive object.
In the high hierarchy (quadrupole approximation) and test particle limit, the angular momentum of the outer orbit is exactly fixed and the coordinate system is chosen such that it is oriented in the direction of the z-axis. The averaged interaction potential turns out to be axisymmetric, and there are no torques along the z-axis and therefore the z-component of the inner binary's (specific) angular momentum $J_{\rm in,z}$ is constant. It is customary to consider the normalized angular momentum, $\mathbf{j}=\mathbf{J}_{\rm in}/J_{\rm in,circ}$, where $J_{\rm in, circ}=\sqrt{G m \ain}$ is the specific angular momentum that the inner binary would have if it were on a circular orbit with semi-major axis $\ain$ and $m$ is the total mass of the inner binaries. The z component of the normalized angular momentum is
constant and given by
\be
j_z=\frac{J_{\rm in,z}}{\sqrt{G m \ain}}=\sqrt{1-e^2}\cos(i),
\ee
where $e$ is the inner binary's eccentricity and $i$ is the mutual inclination between the inner and outer orbits. The quantity $j_z$ is often referred to as the ``Kozai Constant'', which stays constant under the quadrupole approximation of the 
perturbing potential in DA.  When $j_z$ is close to zero, high eccentricities can be obtained.

For moderately hierarchical systems  ($\ain/\aout\lesssim 100$), it has
been recently found that the unaccounted small errors in the approximation employed in the Lidov-Kozai solution may have significant effects on the orbital evolution. These fall into two broad categories -- the long-term evolution of the system
due to higher-order terms in the expansion in $\ain/\aout$ and short-term evolution
due to the non-secular effects:

1) The small contribution of the next order term in the perturbation expansion of $\ain/\aout$ (the octupole term) may accumulate over many Lidov-Kozai cycles and result in significant changes in $j_z$ \citep{ford00, naoz11, katz11, lithwicknaoz11}. In some cases $j_z$ may cross zero so that the orientation of the inner orbit switches between prograde and retrograde with respect to the outer orbit (i.e., orbital flip). When $j_z$ crosses zero, extremely high eccentricities may be obtained.

2a)Within one period of the outer orbit, $j_z$ experiences oscillations which are not described by the Lidov-Kozai approximation \citep{kozaibh2,bodewegg,katzdong}. Very high eccentricities may be achieved if the amplitude of these oscillations is comparable to the magnitude of $j_z$.

2b)
The change in angular momentum may be significant within one period of the inner orbit if the eccentricity is large (i.e. the angular momentum is already close to zero, and see \citet{katzdong,Antonini15} for more discussions). This very short-term change is crucial for head-on collisions as it allows the binary to avoid close passages or grazing encounters (thus avoiding tidal dissipation or tidal disruption or strong GR precession) in the orbits prior to the actual collision \citep{katzdong}.

In this paper we show that the short-term oscillations ignored by the DA approximation can accumulate over time and introduce significant errors in the long-term evolution of moderately hierarchical systems. This occurs when the mass of the perturber is not negligible compared to the central star. Our finding implies that the DA approximation employed in many previous studies of moderately hierarchical systems is inadequate and their results
may need re-examinations. The problem is particularly severe when studying the effect due to the octupole term because it is a long-term effect which is
significant for moderate hierarchy.

The significant error due to breakdown of the DA approximation
had historical importance in celestial mechanics. When the approximation
is used to estimate the apogee precession period of the Moon due to the perturbation of the Sun (precession of the longitude of the periapsis $\varpi$), a value of 18.6 years is obtained, which is about twice the observed value. This led Euler, Delambert and Clairaut to suggest that Newtonian gravity required modification. The problem was eventually solved by Clairaut who corrected the averaging procedure to account for the short-term oscillations \citep[for a historical review, see][]{physicstoday}.

The solution to the lunar problem by Clairaut and its further elaboration made use of the low eccentricity and inclination in the Earth-moon-sun system and therefore cannot be applied to study systems with high inclinations and eccentricities. Corrections for perturbers on circular orbits in the context of irregular moons around giant planets were recently derived by \citet{Cuk04}. We derive the leading-order correction terms (the corrected double-averaging (CDA) equations) that are applicable to any eccentricity of the inner and outer orbit and arbitrary mutual orbital inclination. For simplicity we focus on the test particle case. We show that the CDA equations reduce the long-term error introduced by double-averaging significantly with little extra computing expense.

\section{The double-averaging approximation breaks down over long timescales}\label{sec:DAbreakdown}
\begin{figure}[!htbp]
\includegraphics[scale=0.7]{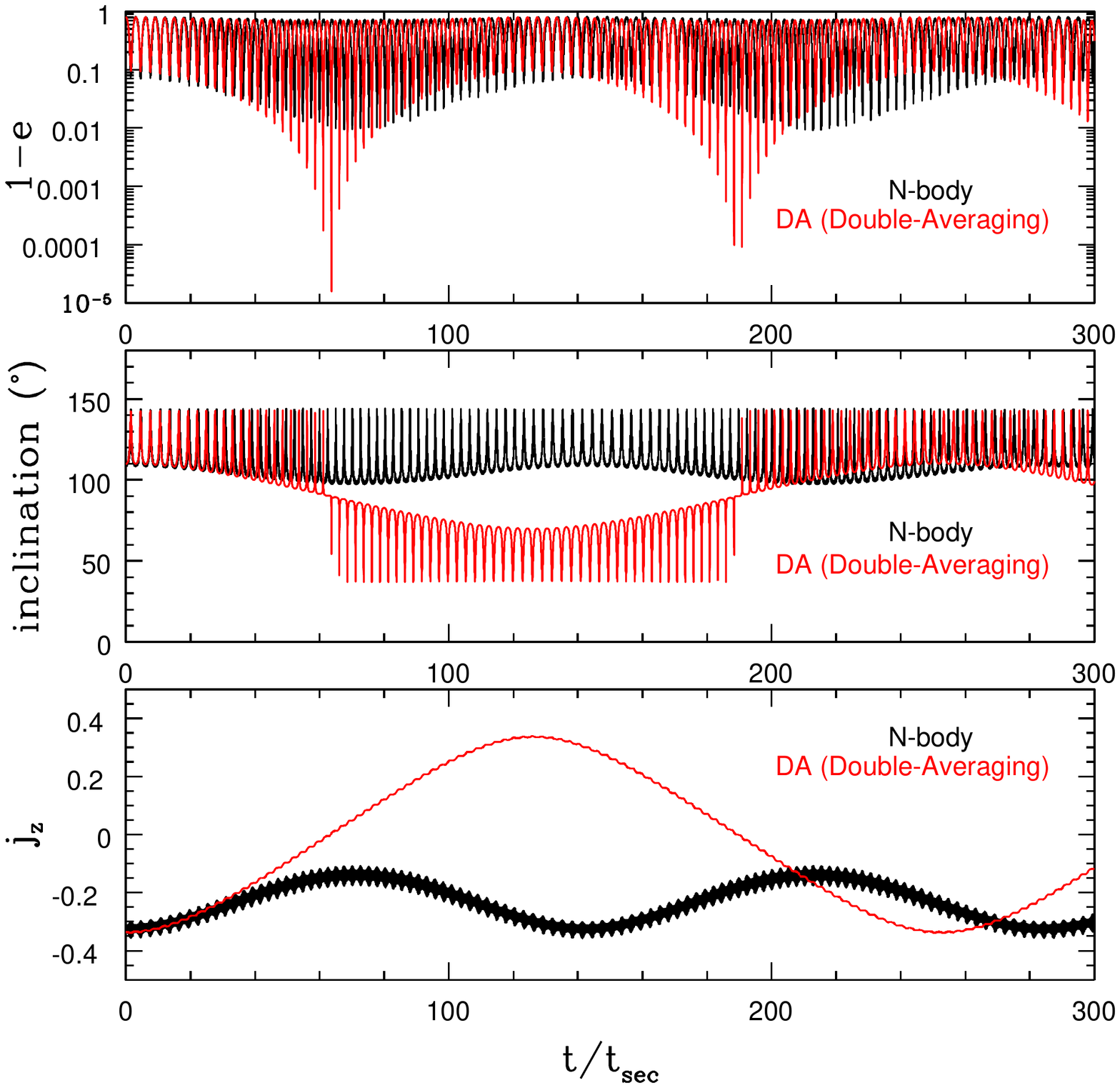}
\caption{Significant difference in the long-term evolution of a three-body system using the exact N-body integration (black) and the approximated double-averaging (DA) integration (red). The system consists of a test particle orbiting a mass $m$ with semi-major axis $a$ and a perturber with the same mass $m$. The outer orbit has semi-major axis $\aout=10a$ and eccentricity $\eout=0.2$. The rest of the parameters are described in the text.
The top, middle and bottom panels show the evolution of $1-e$, inclination  and $j_z=(1-e^2)^{1/2}\cos i$, respectively. The time is normalized to the secular timescale $\tsec$ defined in Eq. \eqref{eq:tsec}.}
\label{fig:eijz}
\end{figure}

We first present an example where the long-term breakdown of the double-averaging approximation is evident. The three-body system in this example consists of a mass $m$ orbited by a test particle with semi-major axis $a$ and a perturber with identical mass $m$ with semi-major axis $\aout=10a$. The z-axis is chosen along the direction of the angular momentum vector of the outer orbit, and the x-axis is in the direction of the eccentricity vector of the outer orbit. Both vectors are fixed in time due to the test particle approximation. The initial eccentricities of both orbits are $e=\eout=0.2$ and the initial inclination is $i=110^\circ$. The initial argument of the pericenter is $\omega=0$ and the longitude of ascending node is $\Omega=\pi$. The evolution of the $e$, inclination and $j_z$ is presented in Figure \ref{fig:eijz} using the accurate, direct N-body (inverse square law) integration. The N-body algorithm applies a Wisdom-Holman \citep{Wisdom91} operator splitting with a high order (8-6-4) coefficient set taken from \citet{Blanes12}, and for more details see descriptions in \citet{katzdong}. The time is shown in units of the secular (Kozai) time scale
\be \label{eq:tsec}
\tsec = \frac{m^{1/2}}{G^{1/2}\mper}\frac{\bout^3}{\ain^{3/2}},
\ee
where $\bout=\aout(1-\eout)^{1/2}$ is the outer semi-minor axis.  As shown in Figure 1., The DA equations fail to reproduce the key characteristics of the long-term evolution calculated from the direct N-body integration. In particular, from the DA integration, the system experiences orbital flips where the inclination crosses $90^\circ$ ($j_z$ crosses $0$) and extremely high eccentricities are obtained at these crossings, but neither such orbital flips nor the accompanying extremely high eccentricities occur from the accurate N-body integration.

\begin{figure}[!htbp]
\includegraphics[scale=0.65]{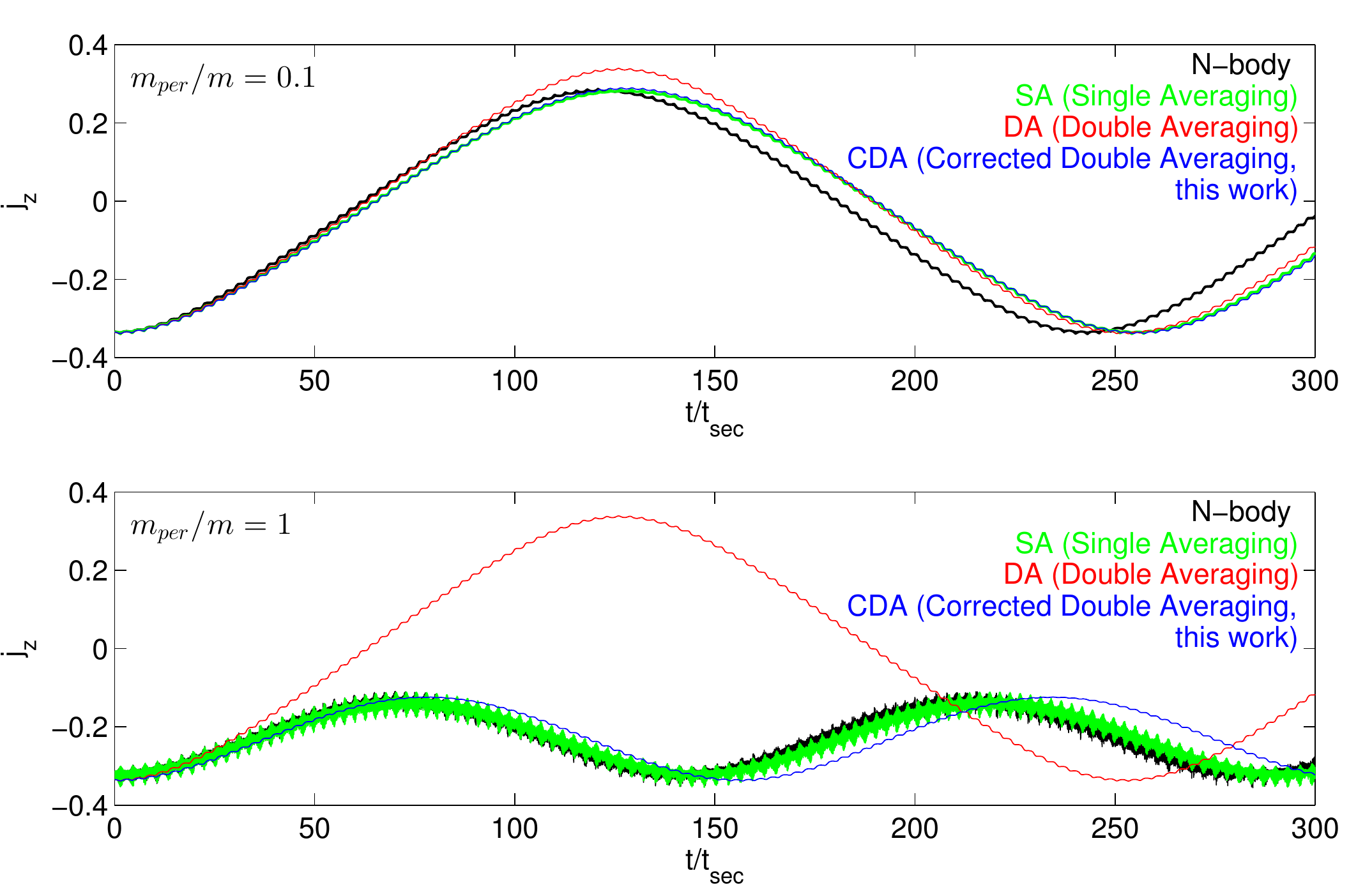}
\caption{Lower panel: 
Long-term evolution of the same system as the bottom panel in Figure \ref{fig:eijz}. Here besides N-body (black) and double-averaging (DA, red), we also include
the results from integrating the single-averaging (SA) equations  (green) and the corrected double-averaging (CDA) equations (blue) derived in \S\ref{sec:CDA}. Upper panel: Long-term evolution of the same system as the lower panel 
except for a small perturber with the ratio between perturber mass and inner binary mass $\mper/m=0.1$. The failure  
of DA in capturing the long-term evolution stems from ignoring 
short-term oscillations on the period of the outer orbit. These short-term oscillations are taken into account by SA, which captures the long-term
behavior of the system. The long-term error of DA is larger for the stronger perturber, which induces short-term oscillations with higher amplitudes. See Figure \ref{fig:jzmper_zoomin} for a closer inspection on the short-term oscillations.}
\label{jzmper}
\end{figure}

\begin{figure}[!htbp]
\plotone{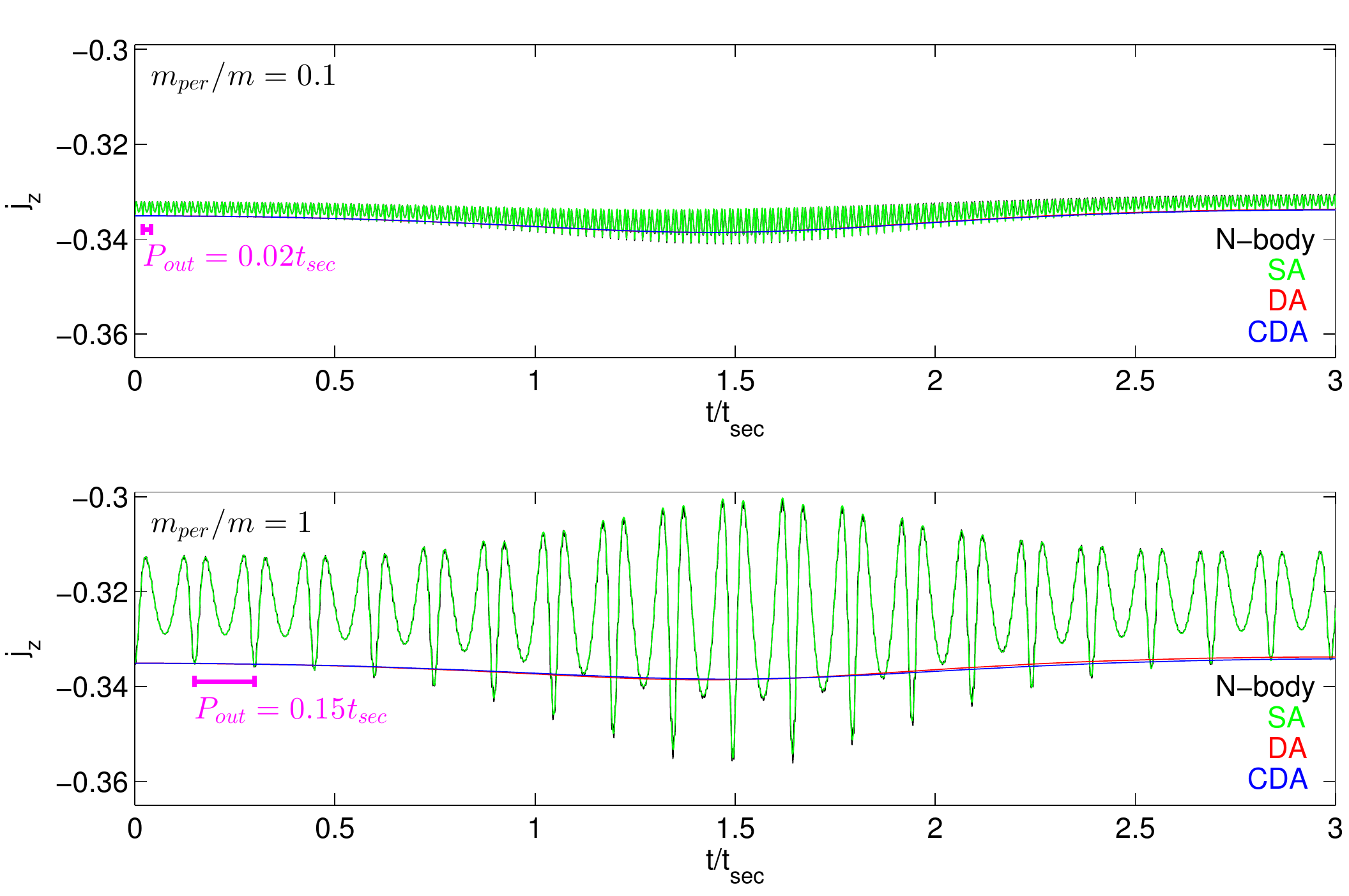}
\caption{Zoomed views on the first Lidov-Kozai cycle of the same systems as shown in Figure \ref{jzmper}, allowing a close inspection on the short-term oscillations that are ignored by DA. The amplitudes for  the short-term oscillations are lower for the less
massive perturber shown in the upper panel.
\label{fig:jzmper_zoomin}}
\end{figure}

This long-term breakdown of the DA approximation stems from ignoring the short-term oscillations in DA. This is demonstrated in Figure \ref{jzmper}. In the lower panel of Figure \ref{jzmper}, we show the results of the N-body and DA integrations, and they are also compared with the single-averaging (SA) approximation where the equations are averaged only over the inner orbit. The potential in the SA approximation is expanded to the octupole term as for the DA equations, therefore the only difference between the SA and DA calculations is whether the outer orbit is averaged (in the DA case ) or not (in the SA case). The results from SA is in good agreement with the N-body integration, implying that the main problem lies in the second averaging over the outer orbit. 

The error in the second averaging is due to the small modulations in the orbital parameters of the inner orbit which occur within each outer orbit. These small oscillations can be clearly seen in the lower panel of Figure \ref{fig:jzmper_zoomin}, which shows the first Lidov-Kozai cycle of the same system as shown in the lower panel of Figure \ref{jzmper}. The upper panels of Figures  \ref{jzmper} and \ref{fig:jzmper_zoomin} show the results of integration for a system with a less massive perturber $\mper=0.1\,m$ and all other initial conditions are identical with those shown in the lower panels. Clearly, the short-term oscillations and long-term errors are much smaller when the perturber is smaller. The effects of the short-term oscillations in Lidov-Kozai cycles were noted before \citep[e.g.][]{kozaibh2,bodewegg,katzdong}. Over a short timescale, If the amplitudes of the short-term oscillations are large enough, they can bring $j_z$ to cross zero, which can lead to high eccentricities \citep[see \S\ref{sec:Introduction} and][]{bodewegg,katzdong,Antonini15}.

Here we study the long-term effects of these short-term oscillations. In \S\ref{sec:CDA}, we calculate these oscillations analytically and then we obtain the ``corrected averaged double-averaged'' (CDA) equations by using these analytical results in the averaging over the outer orbit. The results of the integration of these equations are shown in Figures \ref{jzmper} and \ref{fig:jzmper_zoomin} as blue lines.  As can be seen these equations are in good agreement with the SA approximation and the N-body integration.  In \S\ref{section:statistics} and Appendix \S\ref{Appendix:N-body} several more comparisons between N-body, DA and CDA integrations are performed.

\section{Calculating the short-term Oscillations and Correcting the DA Equations}\label{sec:ShortTermOscCDA}
In this section we derive analytical expressions for the short-term oscillations of $\bfj,\bfe$, and use them to derive corrections to the double-averaging (DA) equations to account for their long-term effects.

Consider an inner binary of two objects with masses $m_2\leq m_1$ and a third body $\mper$. We neglect the changes in the outer orbit which is assumed to be exactly Keplerian with parameters ($\aout$, $\eout$, $\Pout$). This approximation is applicable when considering the short-term oscillations within one outer orbit discussed in \S\ref{sec:ShortTermOsc} and also applicable to studying the long-term effects in two interesting physical cases: $\mper\sim m_1\gg m_2$ (the test particle limit) and $\mper\gg m_1 \sim m_2$ (a binary system with comparable masses orbiting a much more massive object) for which the precession of outer orbit is negligible within the timescale of interest\footnote{Although the change in angular momentum of the outer orbit is small in these cases because it is always much larger than the angular momentum of inner orbit, the Range-Lenz vector may change its direction within the timescale of interest. The precession rate can be estimated as $d\varpi_{out}/d\tau\sim L_{in}/(L_{out}\sqrt{1-\eout^2})$, where $L_{in}=\mu_{in}\sqrt{Gma}$, $\mu_{in}=m_1 m_2/m$ and $L_{out}=\mu_{out}\sqrt{G(m+\mper)\aout}$, $\mu_{out}=m\mper/(m+\mper)$. Our results are applicable for timescales which are much smaller than $(d\varpi_{out}/d\tau)^{-1}\tsec$.}. As in section \S \ref{sec:DAbreakdown}, the z-axis is chosen to be in the direction of the angular momentum vector of the outer orbit and the x-axis pointing along outer orbit's eccentricity vector. We work with a moving coordinate system which is centered on the center of mass of the inner binary, so the position vector of the two inner masses ($m_1$ and $m_2$), $\bfr_1$ and $\bfr_2$, satisfy the relation $m_1\bfr_1+m_2\bfr_2=0$. The perturber's position $\brout(t)$ is confined to the $xy$ plane and is parameterized by the radius $\rout=|\brout|$ and the true anomaly $\fout$ (with $\fout=0$ corresponding to $y=0, x>0$). 

Solving the three-body problem in discussion amounts to finding the trajectory $\bfr(t)=\bfr_2(t)-\bfr_1(t)$ of the inner orbit. The equation of motion for $\bfr$ can be written as
\be
\ddot \bfr = -\nabla_{\bfr}\left[-\frac{Gm}{r}+\Phiper(\bfr,t)\right]
\ee
where
\begin{equation}
m=m_1+m_2
\end{equation}
and
\be
\Phiper(\bfr,t)=-\frac{m}{m_1}\frac{G\mper}{|\brout(t)-\frac{m_1}{m}\bfr|}
-\frac{m}{m_2}\frac{G\mper}{|\brout(t)+\frac{m_2}{m}\bfr|}.
\ee

The perturber is assumed to always be significantly further away compared to the size of inner orbit, and the potential is expanded (using $(1+q^2-2qu)^{-1/2}=\sum_0^\infty q^nP_n(u)$):
\be \label{Eq:Expansion}
\Phiper(\bfr,t)=-\frac{G\mper}{\rout}(\frac{m}{m_1}+\frac{m}{m_2})+\PhiQ+\PhiO
-\frac{G\mper}{\rout}\sum_{n=4}^\infty\tilde m_n\left(\frac{r}{\rout}\right)^nP_n(\cos \theta),
\ee
where $$\cos\theta=\frac{\bfr\cdot \brout}{r\rout}$$,
\be
\tilde m_n=\left(\frac {m_1}{m}\right)^{n-1}+(-1)^{n}\left(\frac{m_2}{m}\right)^{n-1},
\ee
\be
\PhiQ=-\frac{G\mper}{\rout}\left(\frac{r}{\rout}\right)^2P_2(\cos \theta),
\ee
\be
\PhiO=-\frac{m_1-m_2}m\frac{G\mper}{\rout}\left(\frac{r}{\rout}\right)^3P_3(\cos \theta).
\ee
The first term in Eq (\ref{Eq:Expansion}), $-G\mper(m/m_1+m/m_2)/\rout$, does not depend on $\bfr$ and therefore has no affect on $\ddot\bfr$, the second term is the quadrupole potential, the third is the octupole potential, and the last term includes the higher order terms in the potential that are neglected in the analytical derivations presented in this work (and they are of course taken into account for the N-body integrations, which include all terms). 

On short timescales, the trajectory $\bfr(t)$ follows a Keplerian orbit which can  parametrized by the semi-major axis $\ain=-0.5~Gm(\dot \bfr^2/2-Gm/r)^{-1}$, the normalized angular momentum vector $\bfj=\mathbf{\bfr\times \dot \bfr}/\sqrt{Gm\ain}$ and the Runge-Lenz vector $\bfe=\dot \bfr\times(\bfr \times \dot \bfr/(Gm)-\hat r)$ which points in the direction of the pericenter and has a magnitude $|\bfe|=e$. Due to the perturbation, these orbital parameters evolve with time.

All approximations in this paper involve the averaging of the equations of motion over the period of the inner orbit which is the fastest time scale in the problem. The equations of motion take the form \citep[e.g.][]{Tremaine09,Milankovich39}
\be
\frac{d\ain}{dt}= 0,
\ee
and
\begin{align}\label{eq:General}
\frac{d\bfj}{dt}  & =-\frac{1}{\sqrt{Gm\ain}}\cD_{\bfj} \Phi\cr
\frac{d\bfe}{dt} &=-\frac{1}{\sqrt{Gm\ain}}\cD_{\bfe} \Phi~,\cr
\end{align}
where $\cD_{\bfj}$ and $\cD_{\bfe}$ are differential operators defined as follows
\begin{align}\label{eq:DjDedef}
\cD_{\bfj}&=\bfj\times\frac{\partial}{\partial\sbfj} + \bfe\times\frac{\partial}{\partial\bfe}\cr
\cD_{\bfe}&=\bfj\times\frac{\partial}{\partial\bfe} + \bfe\times\frac{\partial}{\partial\sbfj}~,
\end{align}
and $\Phi(\ain,\bfj,\bfe,t)$ is an (appropriately) averaged potential which depends on the approximation involved and may or may not be time dependent. Since $\ain$ is fixed in time (for all of our analytic approximations henceforth), we omit its dependancy.  The different approximations amount to specifying different forms of $\Phi(\bfj,\bfe,t)$.

The most accurate of the approximations considered here is to restrict the averaging to the inner orbit and is obtained by (time) averaging $\Phiper$ for a fixed value of $\brout$ with the inner orbit, $\bfr(t)$ following an exact Keplerian orbit (with parameters $\ain,\bfj,\bfe$). The obtained averaged potential is separated into the expansion terms as in \eqref{Eq:Expansion}, which are averaged separately
\be
\Phi^{SA}=\PhiQ^{SA}+\PhiO^{SA}+...
\ee
In particular, the quadrupole term is given by
\be \label{eq:QSA}
\PhiQ^{SA}(\bfj,\bfe,\brout)=\frac{G\mper \ain^2}{4\rout^3} [-1+6e^2+3(\bfj\cdot\hrout)^2-15(\bfe\cdot\hrout)^2].
\ee

The DA approximation is obtained by averaging the SA equations of motion (Eq. \eqref{eq:General} with $\Phi$=$\Phi^{SA}$) over the outer period by neglecting any changes in the orbital parameters within one outer period. This is equivalent to averaging the potential directly. The resulting quadrupole and octupole terms are given by
\be \label{eq:QDA}
\PhiQ^{DA}(\bfj,\bfe)=\frac34\frac{G\mper \ain^2}{\bout^3} [\frac16+\frac52e_z^2-e^2-\frac12j_z^2],
\ee
and
\be \label{eq:ODA}
\PhiO^{DA}(\bfj,\bfe)=\epoct\frac{75}{64}\frac{G\mper \ain^2}{\bout^3}[2e_zj_xj_z-e_x(\frac15-\frac85e^2+7e_z^2-j_z^2)],
\ee
where $\epoct$ is a small dimensionless number describing the magnitude of $\PhiO$ compared to $\PhiQ$ and is given by
\be\label{eq:epoct}
\epoct=\frac{m_1-m_2}m\frac{\ain}{\aout}\frac{\eout}{1-\eout^2}\sim \frac{\PhiO}{\PhiQ}.
\ee

The Lidov-Kozai approximation (quadrupole, double-averaging) is obtained by using equations \eqref{eq:General} with the approximation $\Phi=\PhiQ^{DA}$ which is expressed in equation \eqref{eq:QDA}. It is straightforward to see that within this approximation, $dj_z/dt=0$.

\subsection{Calulating the Short-Term Oscillations Analytically}\label{sec:ShortTermOsc}
DA ignores the small changes in $\bfj$ and $\bfe$ within the outer orbital period, and we show that such small changes can accumulate and cause a significant error for DA to characterize the long-term evolution of the system. Our task is to calculate such oscillations analytically and redo the averaging to include their effects to correct the DA equations. We will only consider the leading-order (the quadrupole term in Eq. \eqref{eq:QSA}) term of the small oscillations when calculating the oscillations. Note that in the sequent calculations, the corrections of the small oscillations to the leading-order can be incorporated with the higher-order terms (such as octupole) when doing the second averaging over the outer orbit.

When the outer orbit has a high eccentricity, the equations of motion evolve rapidly as function of time when the perturber is in the vicinity of its pericenter due to the fast pericenter passage. It is therefore useful to work with the true anomaly of the outer orbit instead of the time.
The time is related to the outer orbit's true anomaly by
\begin{align}\label{eq:dtdf}
dt=\frac{\rout^2d\fout}{2\pi \aout\bout}\Pout.
\end{align}

We therefore obtain
\begin{align}\label{eq:djedf}
\frac{d\sbfj}{df}&=-\epCDA\,\cD_{\bfj}\phi(\bfj,\bfe,f),\cr
\frac{d\bfe}{df}&=-\epCDA\,\cD_{\bfe}\phi(\bfj,\bfe,f),
\end{align}
where
\begin{equation}\label{eq:epCDA}
\epCDA=\frac{\Pout}{2\pi \tsec}=(\frac{\ain}{\aout})^{3/2}\frac{1}{(1-\eout^2)^{3/2}}\frac{\mper}{[(m+\mper)m]^{1/2}}~,
\end{equation}
\begin{equation}\label{eq:phijef}
\phi(\bfj,\bfe,f)=\frac14(1+\eout \cos\fout)[-1+6e^2+3(\bfj\cdot\hrout)^2-15(\bfe\cdot\hrout)^2]
\end{equation}
and
\begin{align}\label{eq:jedotr}
\bfj\cdot\hrout=j_x \cos\fout+j_y \sin\fout, \cr
\bfe\cdot\hrout= e_x \cos\fout+e_y \sin\fout.
\end{align}

The small parameter $\epCDA$ defined in Eq. \eqref{eq:epCDA} sets the scale of the oscillations within one outer orbit. This is evident from Eq. \eqref{eq:djedf} (noting that $\phi$ is dimensionless and of order unity). Another way to see this is that $\bfj,\bfe$ change by order unity over $\tsec$ and therefore change by order $\Pout/\tsec$ during the time of one outer orbit $\Pout$.

Our approach is to find a  coordinate transformation $(\bar\bfj, \bar\bfe, \fout)\rightarrow(\bfj,\bfe)$ such that the equations of motion of $(\bar\bfj, \bar\bfe)$ will have no dependence on $\fout$. In other words we want to separate $\bfj,\bfe$ into a slow component (independent of $\fout$) and a fast component (dependent of $\fout$). This is done iteratively as an expansion in the small parameter $\epCDA$.
Since $\fout$ and $\fout+2\pi$ are equivalent, the transformation $\bfj(\bar\bfj, \bar\bfe, \fout),\bfe(\bar\bfj, \bar\bfe, \fout)$ depends on $\fout$ periodically and we expand it as a Fourier series in $\fout$,

\begin{align}\label{eq:barje}
\bfj&=\bar\bfj+\epCDA\sum_{l=1} \big[\cos(l\fout)\mathbfcal{J}_l^c(\bar\bfj,\bar\bfe)+\sin(l\fout)\mathbfcal{J}_l^s(\bar\bfj,\bar\bfe)\big],\cr
\bfe&=\bar\bfe+\epCDA \sum_{l=1}
\big[\cos(l\fout)\mathbfcal{E}_l^c(\bar\bfj,\bar\bfe)+\sin(l\fout)\mathbfcal{E}_l^s(\bar\bfj,\bar\bfe)\big],\cr
\end{align}
where $\mathbfcal{J}_l^{c,s}$ and $\mathbfcal{E}_l^{c,s}$ are functions of $\bar\bfe$ and $\bar\bfj$ to be solved for.

It is first useful to expand $\phi$, given in Eq. \eqref{eq:phijef}, as a (finite) fourier series in $f$,
\begin{equation}
\phi=\phi_0+\sum_{l=1}^3\big[\phi_l^c\cos(l\fout)+ \phi_l^s\sin(l\fout)\big].
\end{equation}
We find
\begin{align}\label{eq:phi_lcs}
&\phi_0 =  \frac{1}{8} (1 - 6 e_x^2 - 6 e_y^2 + 9 e_z^2 - 3 j_z^2), \cr
&\phi_2^c = -\frac{3}{8} (5 e_x^2 - 5 e_y^2 - j_x^2 + j_y^2). \cr
&\phi_2^s = -\frac{3}{4} (5 e_x e_y - j_x j_y), \cr
\end{align}
and
\begin{align}\label{eq:phi_lcs1}
&\phi_1^c=\eout(\phi_0+\phi_2^c/2),\cr
&\phi_3^c=\eout\phi_2^c/2, \cr
&\phi_1^s=\phi_3^s=\eout\phi_2^s/2.\cr
\end{align}
The equations of motion can be expanded accordingly
\begin{align}\label{eq:djedfsum}
\frac{d\bfj}{df}&=-\epCDA\bigg(\mathbfcal{J}_{f,0}+
\sum_{l=1}^3\big[\cos(l\fout)\mathbfcal{J}_{f,l}^c+\sin(l\fout) \mathbfcal{J}_{f,l}^s\big]\bigg),\cr
\frac{d\bfe}{df}&=-\epCDA\bigg(\mathbfcal{E}_{f,0}+
\sum_{l=1}^3\big[\cos(l\fout)\mathbfcal{E}_{f,l}^c+\sin(l\fout) \mathbfcal{E}_{f,l}^s\big]\bigg).
\end{align}
where $\mathbfcal{J}_{f,l}^{c},\mathbfcal{J}_{f,l}^{c},\mathbfcal{E}_{f,l}^s,\mathbfcal{E}_{f,l}^c$ are known functions of $\bfj,\bfe$ and are given by
\begin{align}\label{eq:JEflcs}
\mathbfcal{J}_{f,l}^{c,s}&=\cD_{\bfj}\phi_l^{c,s}\cr
\mathbfcal{E}_{f,l}^{c,s}&=\cD_{\bfe}\phi_l^{c,s},
\end{align}
respectively. To be more concrete, we provide here an example of the expression for one of these coefficients:
\begin{equation}
\mathbfcal{J}_{f,2}^{s}=\cD_{\bfj}\phi_2^{c}=\frac{3}{4}\left(-5 e_x e_z+j_x j_z,~~5 e_y e_z-j_y j_z,~~5 e_x^2-5 e_y^2-j_x^2+j_y^2\right).
\end{equation}
Expressions for the rest of the coefficients are given in Eq. \eqref{eq:djedf_coeff}.

Next, $\mathbfcal{J}_{f,l}^{c},\mathbfcal{J}_{f,l}^{c},\mathbfcal{E}_{f,l}^s,\mathbfcal{E}_{f,l}^c$ are used to derive the coefficients $\mathbfcal{J}_{l}^{c},\mathbfcal{J}_{l}^{s},\mathbfcal{E}_{l}^{c},\mathbfcal{E}_{l}^{s}$ in Eq. \eqref{eq:barje}. This is done by working directly with the equations of motion Eq. \eqref{eq:djedfsum}.
By substituting Eq. \eqref{eq:barje} in Eq. \eqref{eq:djedfsum} (on both sides of the equation), saving terms that are up to first order in $\epCDA$ and equating the corresponding coefficients of the cosines and sines, we obtain
\begin{align}\label{eq:jel1}
\mathbfcal{J}_l^c&=\frac{\mathbfcal{J}_{f,l}^s}{l}+O(\epCDA), ~~\mathbfcal{J}_l^s=-\frac{\mathbfcal{J}_{f,l}^c}{l}+O(\epCDA)\cr
\mathbfcal{E}_l^c&=\frac{\mathbfcal{E}_{f,l}^s}{l}+O(\epCDA),
~~\mathbfcal{E}_l^s=-\frac{\mathbfcal{E}_{f,l}^c}{l}+O(\epCDA)
\end{align}
all to be evaluated at $\bfj,\bfe=\bar\bfj,\bar \bfe$.
The coordinate transformation thus reads:
\begin{align}\label{eq:jebar1}
\bfj&=\bar\bfj+\epCDA\sum_{l=1}^3\bigg[\frac{\cos(l\fout)}{l}\mathbfcal{J}_{f,l}^s(\bar \bfj,\bar \bfe)-\frac{\sin(l\fout)}{l} \mathbfcal{J}_{f,l}^c(\bar \bfj,\bar \bfe)\bigg]+O(\epCDA^2),\cr
\bfe&=\bar\bfe+\epCDA\sum_{l=1}^3\bigg[\frac{\cos(l\fout)}{l}\mathbfcal{E}_{f,l}^s(\bar \bfj,\bar \bfe)-\frac{\sin(l\fout)}{l} \mathbfcal{E}_{f,l}^c(\bar \bfj,\bar \bfe)\bigg]+O(\epCDA^2).
\end{align}
where by $\mathbfcal{J}_{f,l}^s(\bar \bfj,\bar \bfe)$ we mean the function $\mathbfcal{J}_{f,l}^s$ evaluated at $\bfj=\bar \bfj, \bfe=\bar\bfe$.
In particular, to first order in $\epCDA$, the transformation of $j_z$ is
\begin{align}\label{eq:jzbar1}
j_{z}=\bar j_z-&\epCDA C[\eout\cos(\fout)+\cos(2\fout)+\eout\cos(3\fout)/3]\cr
+&\epCDA S[\eout\sin(\fout)+\sin(2\fout)+\eout\sin(3\fout)/3],
\end{align}
where
\begin{align}\label{eq:jzbarCS}
&C=\frac38(5\bar e_x^2-5\bar e_y^2-\bar j_x^2+\bar j_y^2)\cr
&S=\frac34(-5\bar e_x\bar e_y+\bar j_x\bar j_y).
\end{align}

\begin{figure}[!htbp]
\plotone{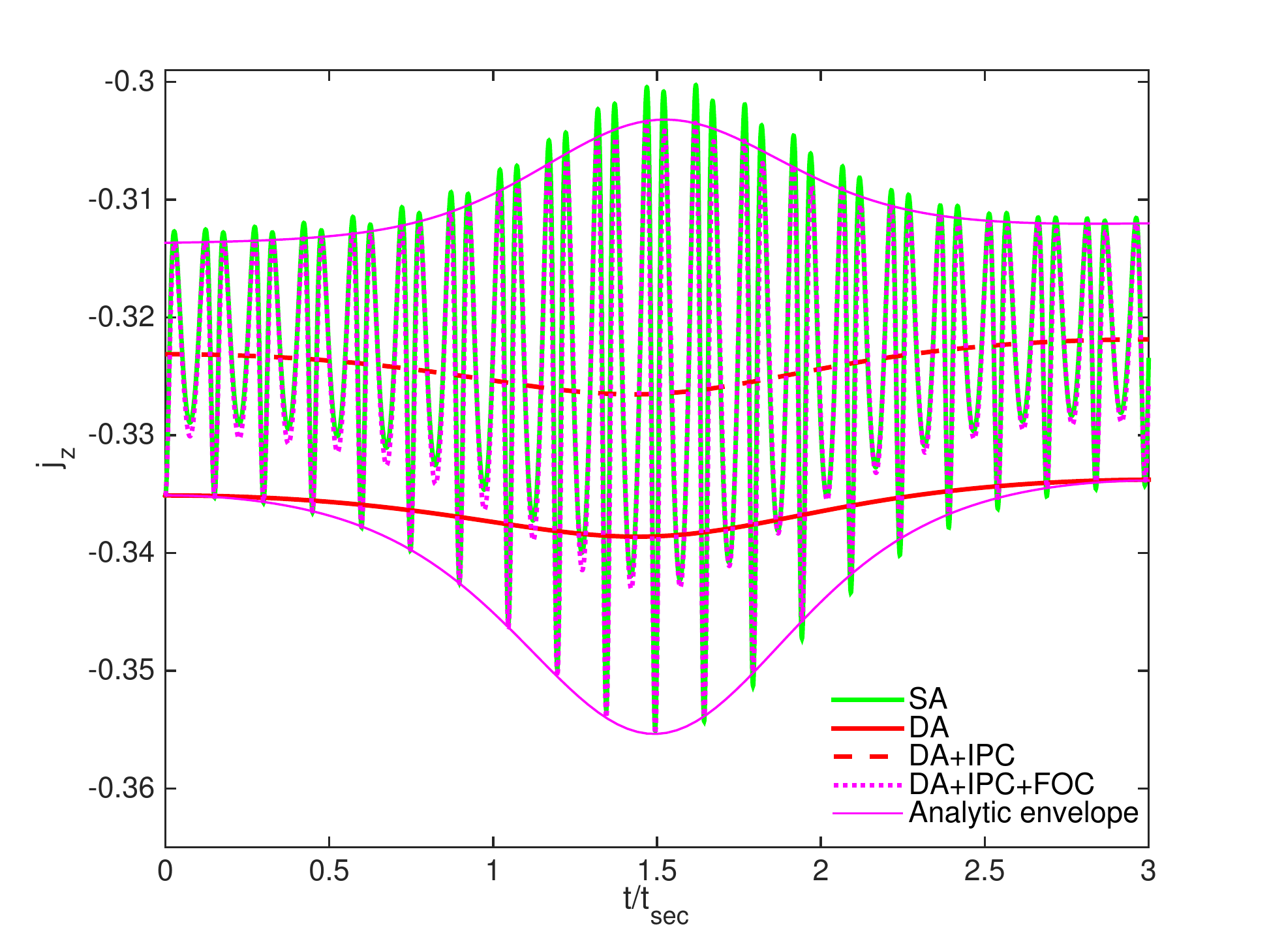}
\caption{Initial Phase Correction (IPC) and Fast Oscillation Component (FOC). The results are for the same system as shown in the lower panel of Figure \ref{fig:jzmper_zoomin}. The results of a single-averaging approximation (SA) calculation are shown in green, while the double-averaging (DA) solution is shown in solid red. DA with IPC is shown in dashed red. For IPC, the initial conditions for $\bar\bfj$ and $\bar\bfe$ are found using the initial value of $\fout$ and Eq. \eqref{eq:jebar1}, and are later evolved using the DA approximation. We also show the FOC of $j_z$ using $\fout$ and equation \eqref{eq:jzbar1}, shown in dotted magenta. The envelope of the oscillations is calculated using Eq. \eqref{eq:jzenv} and shown as solid magenta.
\label{jzmper_zoomin_OSC}}
\end{figure}

To this order, the equations of motion for $\bar \bfj, \bar \bfe$ are
\begin{align}\label{eq:djedf0}
\frac{d\bar \bfj}{df}&=-\epCDA\mathbfcal{J}_{f,0}=-\epCDA\,\cD_{\bar\bfj}\phi_0,\cr
\frac{d\bar \bfe}{df}&=-\epCDA\mathbfcal{E}_{f,0}=-\epCDA\,\cD_{\bar\bfe}\phi_0
\end{align}
where $\mathbfcal{J}_{f,0},\mathbfcal{E}_{f,0},\phi_0$ should be evaluated at $\bfe=\bar\bfe, \bfj=\bar\bfj$. Eq. \eqref{eq:djedf0} is equivalent to the (quadrupole) DA equation.

Equation \eqref{eq:jebar1} captures the short-term oscillations discussed in \S\ref{sec:DAbreakdown}. To demonstrate this, the resulting oscillations in $j_z$ are compared to the results of the SA approximation in Figure \ref{jzmper_zoomin_OSC}, which shows the same system as the bottom panel of Figure \ref{fig:jzmper_zoomin}. We first calculate $\bar\bfj,\bar\bfe$ by integrating the DA equations with the initial value of $\bar\bfj,\bar\bfe$ obtained from the initial value $\bfe,\bfj$, $\fout$ by solving (numerically) Eq. \eqref{eq:djedf0}. The resulting evolution of $\bar j_z$ is shown in dashed red. The only difference with respect to the DA approximation (shown in the Figure in solid red) is the correction in the initial conditions which is henceforth denoted as Initial Phase Correction (IPC). As can be seen in the Figure, $\bar\bfj,\bar\bfe$ represent the middle of the oscillations between the two extremes. The oscillations are then calculated using Eq. \eqref{eq:jebar1} [or equivalently using Eqs. \eqref{eq:jzbar1},\eqref{eq:jzbarCS}] and shown in dotted magenta, which are in good agreement with the SA approximation (shown in solid green).

One of the interesting questions in the long-term evolution of $j_z$ is wether or not it can cross $0$. It is therefore useful to express the envelope of the oscillations in $j_z$. Using Eq. \eqref{eq:jzbar1}, it is straightforward to show that the maximal and minimal values of $j_z$ for $0<\fout<2\pi$ are given by
\begin{align}\label{eq:jzenv}
j_{z,\max}=\bar j_z +\epCDA\sqrt{C^2+S^2}\left(1+\frac{2\sqrt{2}}{3}\eout\sqrt{1-\frac{C}{\sqrt{C^2+S^2}}}\right)\cr
j_{z,\min}=\bar j_z -\epCDA\sqrt{C^2+S^2}\left(1+\frac{2\sqrt{2}}{3}\eout\sqrt{1+\frac{C}{\sqrt{C^2+S^2}}}\right),\cr
\end{align}
where $C,S$ are given in Eq. \eqref{eq:jzbarCS}.
The envelope obtained by equation \eqref{eq:jzenv} is shown in Figure \ref{jzmper_zoomin_OSC} as solid magenta lines, which agree with the extremes of the short-term oscillations.

\subsection{Correcting the double-averaging equations}\label{sec:CDA}

\begin{figure}[!htbp]
\includegraphics[scale=0.7]{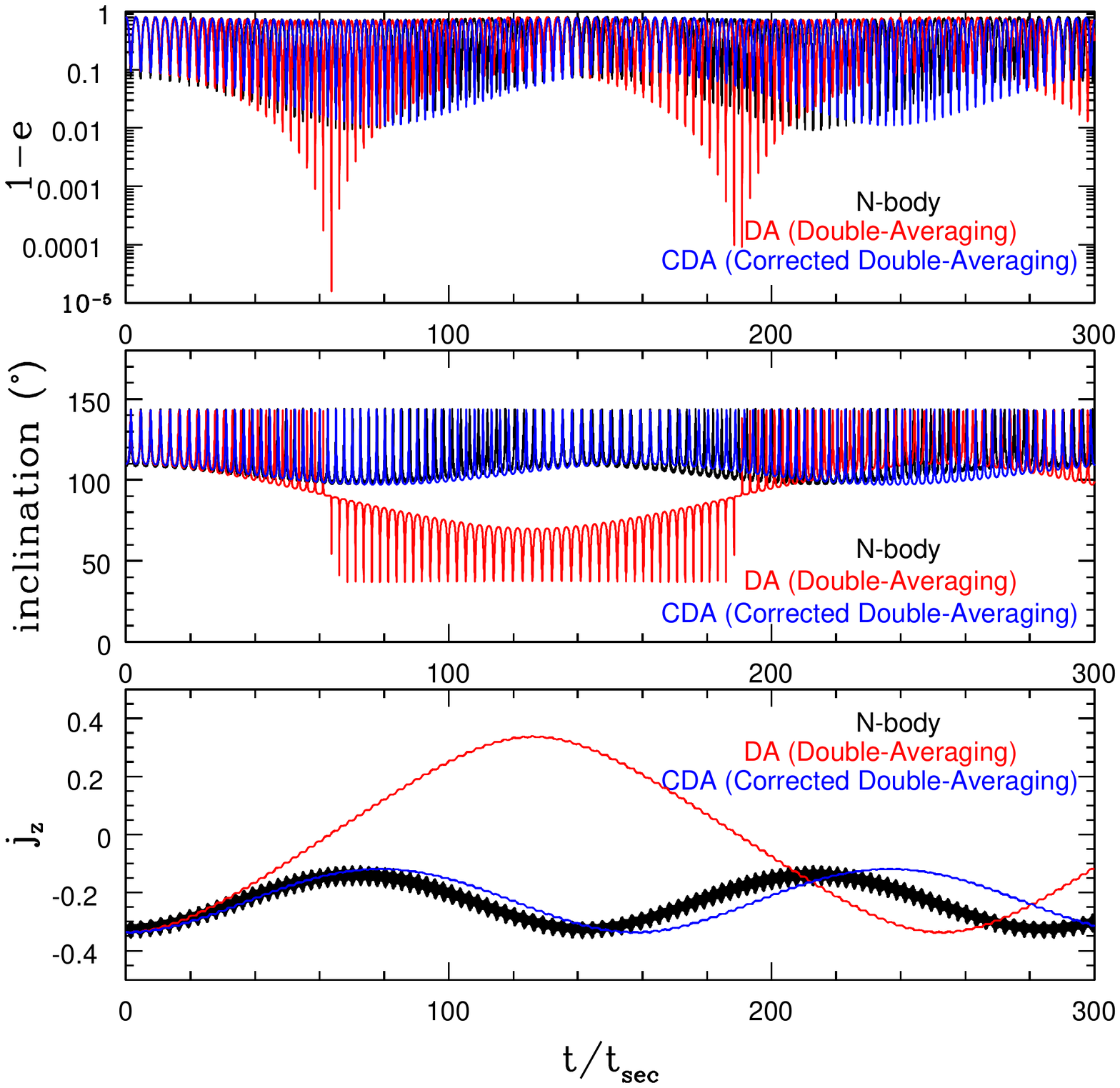}
\label{cda_eijz}
\caption{Comparison of the accurate N-body (black), double-averaging (DA, red) and corrected double-averaging (CDA, blue) calculations for the same system and parameters as in Figure \ref{fig:eijz}. The results of the comparisons show that 
that CDA captures the long-term characteristics in the evolutions of eccentricity, inclination and $j_z$ that are not captured by DA.}
\end{figure}

Next, we substitute Eq. \eqref{eq:barje} in Eq. \eqref{eq:djedfsum}, saving terms that are up to second order in $\epCDA$, using the leading terms of $\mathbfcal{J}_l^{c,s},\mathbfcal{E}_l^{c,s}$ obtained in Eq. \eqref{eq:jel1}. In this second iteration we are not interested in the updated expressions for $\mathbfcal{J}_l^{c,s},\mathbfcal{E}_l^{c,s}$ but rather in the remaining part of the equation that does not depend on $f$, namely  $\mathbfcal{J}_{f,0},\mathbfcal{E}_{f,0}$. This can be found by substituting the expression for $\bfj,\bfe$ in Eq. \eqref{eq:jebar1} in the right hand side of equation Eq. \eqref{eq:djedfsum} and averaging over $f$. For example, the averaged equation for $\bar e_x$ is

\begin{align}\label{eq:dbarexdf}
\big(\frac{d\bar e_x}{df}\big)_{\epCDA}=&
\epCDA\left(\frac{27}{64}(6\bar e_z \bar j_y \bar j_z+\bar e_y(\frac{1}{3}+8\bar e_x^2+8\bar e_y^2+3\bar e_z^2-17\bar j_z^2)\right)\cr
&+\epCDA\eout^2
\left(\frac{9}{64}(14\bar e_z \bar j_y \bar j_z+\bar e_y(\frac{35}{3}+10\bar e_x^2+5\bar e_z^2-10\bar j_x^2-32\bar j_y^2-35\bar j_z^2)\right).
\end{align}

Since we are after the long-term evolution of $\bar\bfj,\bar\bfe$, solving the equations as a function of $f$ or as a function of time is equivalent. It is convenient to measure time using the Lidov-Kozai timescale $\tsec$,
\begin{equation}\label{eq:deftau}
\tau=t/\tsec.
\end{equation}
Using equations Eq. \eqref{eq:tsec} and \eqref{eq:dtdf}, the average rate of change of $f$ and $\tau$ are related by
\begin{equation}\label{eq:dtaudf}
\langle df \rangle = \epCDA^{-1} d\tau.
\end{equation}
The resulting equations using $\tau$ are given in Eq. \eqref{eq:fullCDA}. For completeness we also provide the the doubly-averaged equations in Eq. \eqref{eq:fullDA}. Eq. \eqref{eq:fullCDAterms}, which is the sum of Eq. \eqref{eq:fullCDA} and Eq. \eqref{eq:fullDA} is termed  ``corrected double-averaging''  (CDA) equation in this paper.

If we add the following additional potential to the DA potential [given in Eq. \eqref{eq:QDA}], the equations of motion can be derived using Eq. \eqref{eq:General}:
\begin{align}\label{eq:CDAPotential}
\Phi_{CDA}=-\epCDA\frac{G\mper \ain^2}{\bout^3}&\Big(\frac{27}{64}j_z[(1-j_z^2)/3+8e^2-5e_z^2]+\cr
		&+\frac{3\eout^2}{64}[e_z(10j_xe_x-50j_ye_y)+j_z(5j_x^2-j_y^2+65e_x^2+35e_y^2)]\Big).
\end{align}
The potential in Eq. \eqref{eq:CDAPotential}, reduces to the potential derived by \citet{Cuk04} in the limiting case that they considered of $\eout=0$ and $\mper\gg m$, if we replace the term $\sin^2\omega$ (which appears in $e_z^2$ in Eq.\eqref{eq:CDAPotential}) with its averaged value $1/2$. It is not clear to us why the potential we derived depends on $\omega$ while the potential derived by \citet{Cuk04} does not. 

We show the results by integrating the CDA equations in Figure \ref{cda_eijz}, and 
the three-body system is the same as shown in Figure \ref{fig:eijz}. As can be seen, the CDA equations manage to correct most of the long-term error present in the DA equations.  In Sections \S\ref{section:statistics} and \S\ref{Appendix:N-body} several more comparisons between N-body, DA and CDA integrations are performed. 

\subsection{Implementation - Using the Equations in Computer Codes}
\label{sec:Implementation}
A {\it Matlab} implementation of the equations presented in \S\ref{sec:ShortTermOsc} and \S\ref{sec:CDA} is provided in the supplementary  text files. The digital forms of the equations therein can be easily adapted to other programming languages. We provide below brief descriptions on what these codes do and how they can be used. The codes are also annotated with comments. 

\subsubsection{Long-Term Evolution using the CDA Equations}
The long-term evolution of the inner orbit can be calculated using equations \eqref{eq:fullCDAterms}-\eqref{eq:fullDA}. The corrected double-averaging approximation (CDA) equations are implemented in the file $CDA\_Derivative.m$ which includes a function that calculates $d\bfe/d\tau$ and $d\bfj/d\tau$. When setting $\epCDA=0$, the function provides the (``un-corrected'') double-averaging approximation (DA) equations. We supply the parameters used in the example shown in Figure \ref{jzmper_zoomin_OSC} and the bottom panel of Figure \ref{fig:jzmper_zoomin} in $Integration\_Example\_System.m$, and this
script can be used to perform the integration to obtain the long-term evolution 
of the system shown in those Figures. 
When integrating the DA or the CDA equations, a slightly better solution 
with the Initial Phase Correction (IPC) can be obtained by converting the initial conditions of $\bfj,\bfe$ to the corresponding values of $\bar \bfj, \bar \bfe$ using Eq.\eqref{eq:jebar1}. 
IPC (see Figure \ref{jzmper_zoomin_OSC} and corresponding discussions in the text) can be done only if the initial value of the true anomaly $\fout$ is speficied. Parameters in the beginning of the file $Integration\_Example\_System.m$ allow the users
to choose to turn on or off IPC and whether to carry out calculations using the DA or CDA equations.

\subsubsection{Short-Term Oscillations}
The oscillations of $\bfj,\bfe$ within an outer orbital period can be calculated analytically using equations \eqref{eq:jebar1}, \eqref{eq:djedf_coeff} and \eqref{eq:phi_lcsA} for  given values of $\bar \bfj,~\bar \bfe$, $\eout$, $\epCDA$ [defined in Eq. \eqref{eq:epCDA}] and $\fout$. 
This is implemented in the file $Quadrupole\_Pout\_Oscillation.m$. This result is applicable to first order to any configuration and set of masses (i.e., it is not limited to the test particle approximation). 
For calculating the oscillations of $j_z$, Eqs. \eqref{eq:jzbar1}-\eqref{eq:jzbarCS} can be used, and they are implemented in $Quadrupole\_Pout\_Oscillation\_jz.m$. The envelope of the oscillations of $j_z$ is given by Eq. \eqref{eq:jzenv} and implemented in $Quadrupole\_Pout\_Oscillation\_jz\_maxmin.m$. The envelope is useful for long-term calculations in which it is desirable to avoid the relatively large number of time steps required to resolve the outer orbits. The short-term oscillations for the example system shown in Figure \ref{jzmper_zoomin_OSC} can be calculated with the script  $Integration\_Example\_System.m$ for the initial stages of the evolution.  To do this, the values of $\bar\bfj$,$\bar\bfe$ are calculated in the script by performing a long-term integration and are later used when calling the function $Quadrupole\_Pout\_Oscillation\_jz.m$ to find the short-term oscillations.

\section{Comparison of N-Body, DA and CDA }
\label{section:statistics}
In this section, we compare N-body, DA and CDA using large sets of initial conditions.   
We focus on comparing the conditions of orbital flips (i.e., whether 
$j_z$ crosses zero).
The DA and CDA equations are integrated using the fourth-order Runge-Kutta method with a fixed time step $dt = 0.05\,\tsec$. Each run is stopped when reaching $t_{max}=10 \epoct^{-1}\tsec$ to make it sufficiently long for the octupole term to have a considerable effect. The N-body calculations are performed using the Wisdom-Holman splitting with adaptive time step described in \citet{katzdong}. The presented examples are to illustrate some significant differences between DA and CDA for several large ensembles of systems. We stress that these examples are far from a complete survey of the parameter space, which is beyond the scope of this paper.   

\begin{figure}[h!]
\plotone{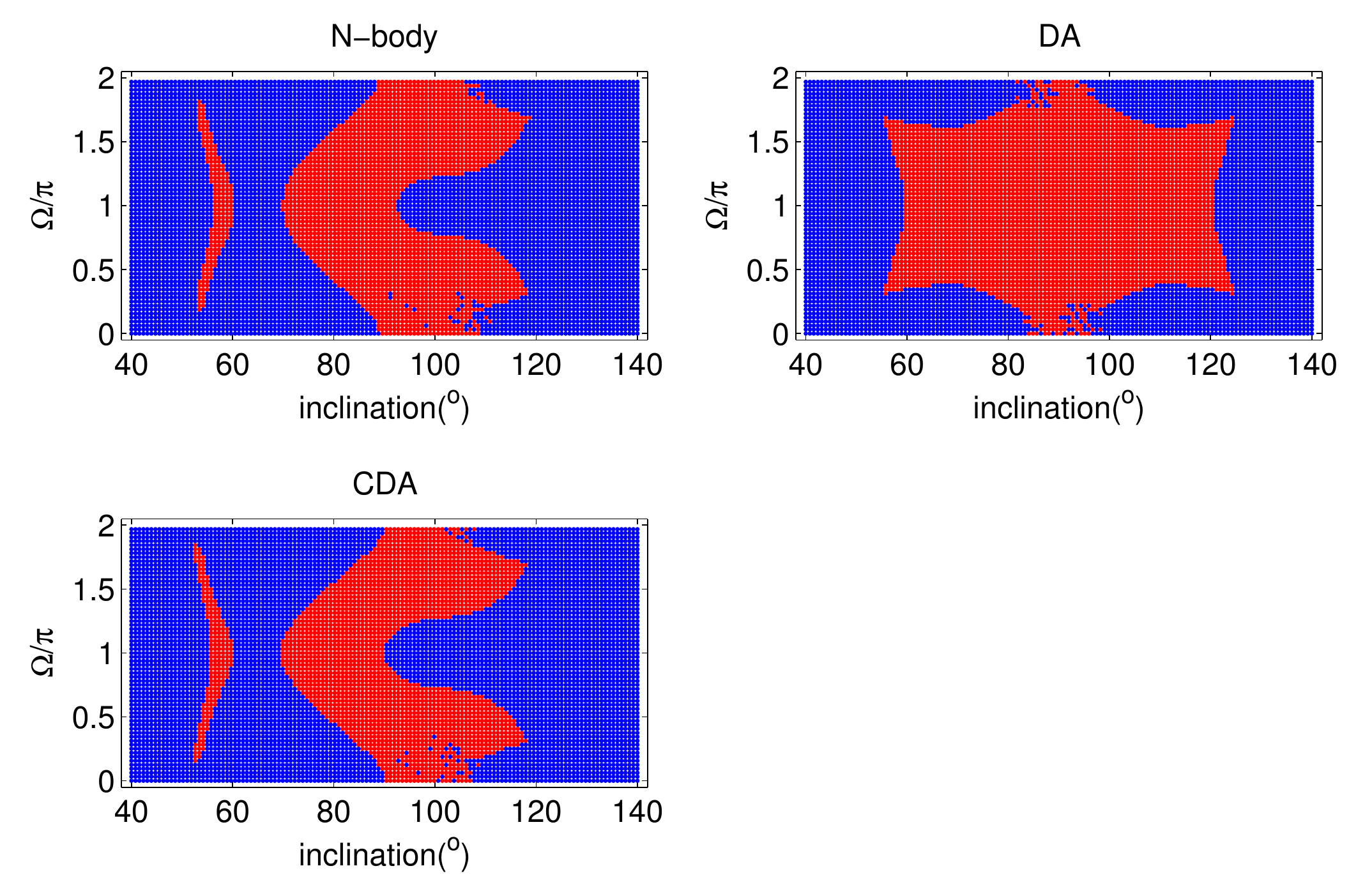}
\caption{Parameter space of orbital flips resulting from the N-body, DA and CDA integrations. Initial orbital parameters: $\mper/m=1$, $\omega=0$, $\eout=0.2$, $\aout/a=10$, $e=0.2$. The initial values of inclination and $\Omega$ are scanned. For the N-body integrations, the true anomaly of both outer orbit and inner orbit are initially set to 0. Integrations in which flips occur (i.e., $j_z$ crossed zero) are shown as red points while those without are shown in blue. As can be seen, the DA
calculations have significantly different results from N-body while CDA corrects
most of the errors in DA.}
\label{flip_aper_10}
\end{figure}

Figure \ref{flip_aper_10} shows the results of comparing  N-body, DA and CDA integrations with 
changing orbital orientations. The integrations share the 
following parameters [$\mper/m=1,\aout/a=10$, $\eout=0.2$] and initially $[\omega=0,~e=0.2]$. The initial values of inclination and $\Omega$ are scanned. The choice of the initial true anomaly has insignificant effects here and is set to be 0 in the N-body calculations. For each simulation we record whether a flip occurs or not (i.e., whether $j_z$ crosses 0) during the entire run, and the results (flip/non-flip) are denoted as dots in different colors (red=flip, blue= no-flip). The IPC and FOC corrections (see \S\ref{sec:ShortTermOsc}) are not taken into account in the DA or CDA integrations. As can be seen from comparing with the N-body results (upper left), the 
DA integrations (upper right) fail significantly in capturing the bulk of 
the parameter space where orbital flips occur. CDA results (lower left) show 
good consistency with the N-body results.  Similar comparisons for other sets of $\aout/a$, $e$ and $\eout$ are shown in Figures \ref{flip_aper_5}-\ref{flip_aper_30_eper_0d8}.

\begin{figure}[h!]
\plotone{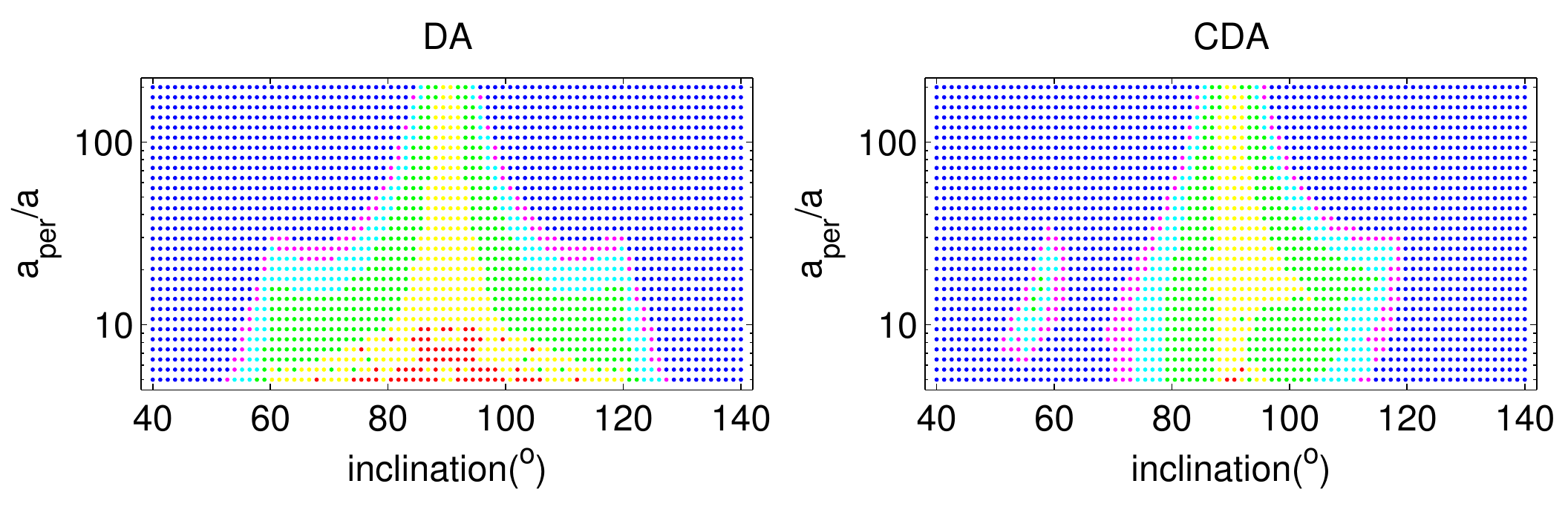}
\caption{Orbital flip fraction as a function of inclinations and hierarchy (i.e., 
outer and inner semi-major axis ratios). The flip fraction is calculated for each combination of the inclination and semi-major axis ratio $\aout/a$ by performing 20 integrations with a range of $\Omega$ values between $0$ and $2\,\pi$ in steps of $0.1\,\pi$. All integrations have the initial parameters $\omega=0$, $\eout=0.2$, $e=0.2$ and $\mper/m=1$. The results for integrations using the DA equations are shown in the left panel and those using the CDA equations in the right panel. The fraction (among the 20 runs) is illustrated by using different color -- blue=0, magenta=(0,0.25], cyan=(0.25,0.5], green=(0.5,0.75], yellow=(0.75,1), red=1.}
\label{flipfrac_e_0d2}
\end{figure}

\begin{figure}[ht!]
\plotone{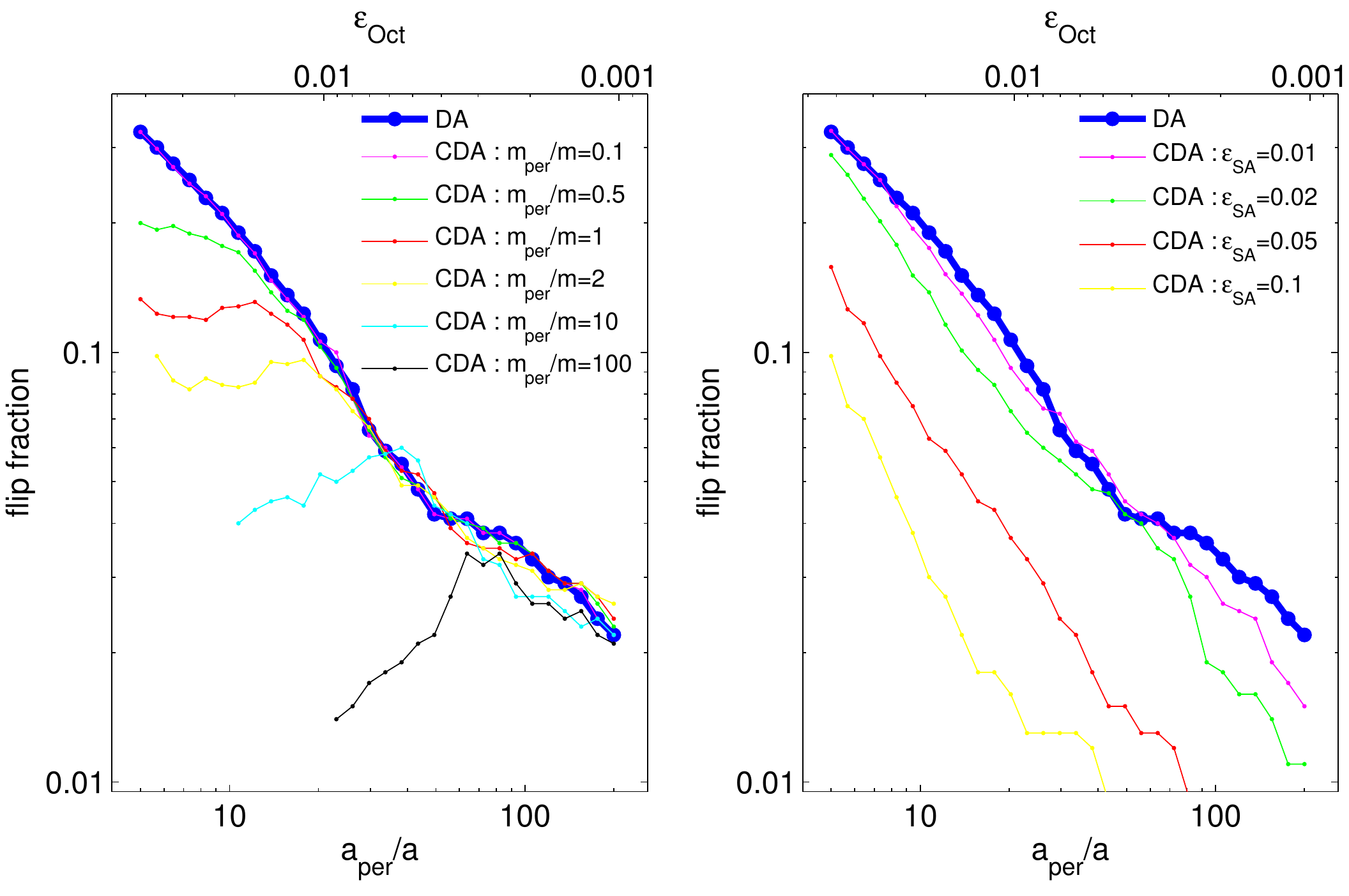}
\caption{Flip probability for isotropic orientations as a function hierarchy (the outer and inner semi-major axis ratios) for different outer-to-inner mass ratios.
The flip fraction is calculated for each combination of the mass ratio $\mper/m$ and semi-major axis ratio $\aout/a$ by performing 1000 integrations with orientations randomly chosen from an isotropic distribution. All integrations have the initial eccentricities $\eout=e=0.2$. Under the test particle approximation, the DA results are the same for all mass ratios (shown in blue) while the CDA results are presented for different mass ratios in different colors (see the legends and descriptions below). Left Panel: The flip probability as a function of  $\aout/a$ for 6 fixed values of $\mper/m$ ranging from 0.1 to 100 shown in different colors (see the legend for the values and corresponding colors). Right Panel: The flip probability as a function of $\aout/a$ for 4 fixed values of the expansion parameter $\epCDA$ (Eq. \eqref{eq:epCDA}) ranging from 0.01 to 0.1 shown in different colors (see the legend for the values and corresponding colors). For each combination of $\aout/a$ and $\epCDA$, the mass ratio $\mper/m$ is calculated using Eq. \eqref{eq:epCDA}.  As can be seen, significant differences between the DA and CDA calculations occur when $\epCDA\gtrsim \epoct$, where $\epoct$ is the coefficient characterizing the strength of the octupole, shown in the top x-axis. 
term  (Eq. \eqref{eq:epoct}).}
\label{random}
\end{figure}

In Figure \ref{flipfrac_e_0d2}, the fraction of flips when scanning over $\Omega$ and fixing $e=\eout=2$ is shown as a function of inclination and the ratio of the semi-major axises. For each combination of the inclination and $\aout/a$,  20 runs are performed with $\Omega$ uniformly distributed between 0 to $2\pi$ (step of $0.1\pi$) to calculate the flip fraction. Different colors are used to denote different ranges of flip fractions with a step size of 0.25 (see caption). As can be seen, for the chosen masses and initial conditions, the flip fractions are similar between DA and CDA for $\aout/a \gtrsim 30$ while 
there are substantial differences at smaller semi-major axis ratios.

\begin{figure}
\plotone{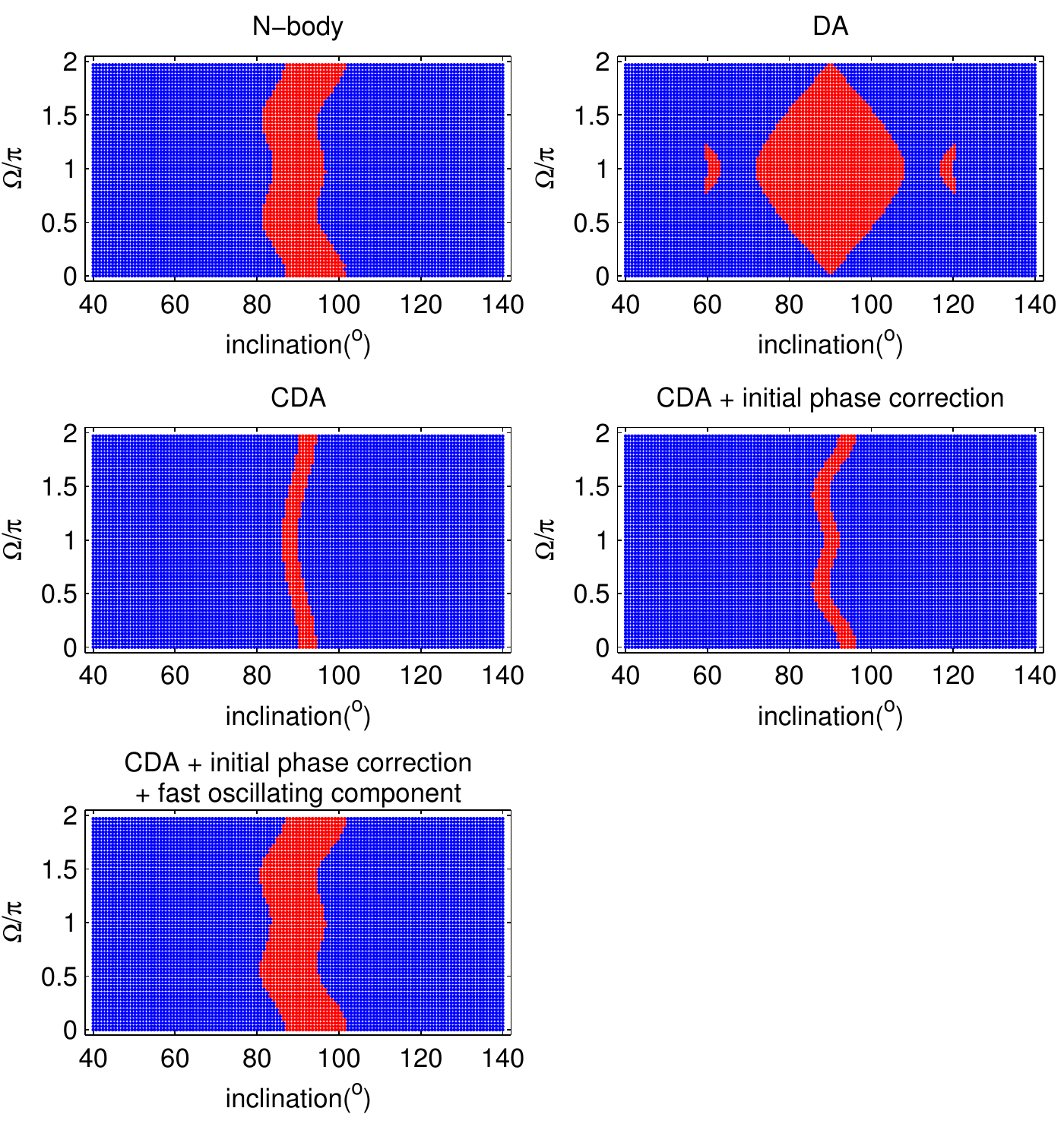}
\caption{
Parameter space for orbital flips in the case of a massive perturber, $\mper/m=100$. The results of  N-body, DA, CDA, CDA + ``Initial phase correction'' (IPC), and CDA+``Fast Oscilsating Component'' (FOC, see Figure \ref{jzmper_zoomin_OSC} and related discussion in section \S\ref{sec:ShortTermOsc}) are shown. Input orbital parameters: $\omega=0$, $\eout=0.2$, $\aout/a=30$, $e=0.2$, $\fout=0$ (for the N-body code the true anomaly of the inner orbit is also 0).  As in Figure \ref{flip_aper_10}, the results are shown as a function of the initial inclination and $\Omega$ and flip/non-flips are marked by red/blue points. As can be seen, for such massive perturbers the IPC and FOC corrections are required to capture the flips correctly.
}
\label{flip_IPC_FOC}
\end{figure}

One generic feature that can be seen is a difference in the symmetry properties of the different approximations. In the DA approximation, the equations do not depend on the direction in which the perturber moves along the outer orbit (prograde or retrograde). The DA equations are therefore invariant under the transformation $\bfj_{\out}\rightarrow -\bfj_{\out}$ which results in a change in our coordinate system $(\hat y, \hat z)\rightarrow (-\hat y,-\hat z)$ and therefore $(\text{inclination}, \Omega,\omega)$ to $(\pi - \text{inclination}, - \pi-\Omega,\omega+\pi)$. Combining this with the mirror symmetry with respect to the $xy$ plane of the N-body equations $z,v_z->-z,-v_z$, which results in ($e_z,j_x,j_y\rightarrow -e_z,-j_x,-j_y$). This mirror symmetry does not affect the coordinate system or inclination but results in $\Omega,\omega\rightarrow \Omega+\pi,\omega+\pi$. Combining the symmetries we obtain that the DA equations (to all orders in the expansion in $a_{\rm in}/a_{\rm out}$) are symmetric with respect to $(\text{inclination}, \Omega,\omega)$ to $(\pi - \text{inclination}, -\Omega,\omega)$. When sampling  over all values of $\Omega$, this implies a symmetry of the form $\text{inclination}\rightarrow \pi-  \text{inclination}$.
 As can be seen in Figures \ref{flip_aper_10},\ref{flipfrac_e_0d2}, while the results of the DA approximation respect this symmetry, the results of the N-body and CDA calculations violate this symmetry significantly. In particular note an interesting ``island'' in the right panel of Figure \ref{flipfrac_e_0d2} at inclination $\sim50^\circ$ and $\aout/a\sim 10$ where flips occur in the CDA calculations at relatively low inclinations with no counterparts on the retrograde region. No ``islands'' exist in the DA results.

Finally, the flip fractions for isotropic distributions as a function of $\aout/a$ and $\mper/m$ are shown in Figure \ref{random}. For each set of  we perform Monte-Carlo simulations using the test-particle DA and CDA equations for a wide range of values of $\aout/a$ and $\mper/m$. For each combination of $\aout/a$ and $\mper/m$, the flip probability is calculated using 1000 simulations with randomly chosen orientations drawn from an isotropic distribution  (the cosine of inclination, $\Omega$ and $\omega$ follow uniform distributions). The initial eccentricities are fixed to $e=\eout=0.2$. 
The resulting flip probabilities as a function of $\aout/a$ for a number of perturber masses are shown in the left panel of Figure \ref{random}. As can be seen in the Figure, the DA flip probabilities do not depend on the mass ratio $\mper/m$ while CDA probabilities do. In general the long-term correction in included in the CDA equations decreases the flip-probability.

A rough criterion for the importance of the correction to the DA equations can be obtained by comparing the dimensionless coefficients $\epoct$ and $\epCDA$ (see Eqs. \eqref{eq:epoct}, \eqref{eq:epCDA} in section \S\ref{sec:ShortTermOscCDA}) which represent the relative magnitude of the octuple terms (responsible for the flip) and the new correction terms (as compared to the quadruple terms). For convenience we provide them here:
\begin{align}
\epoct&=\frac{m_1-m_2}{m}\frac{a}{\aout}\frac{\eout}{1-\eout^2}~,\nonumber\\
\epCDA&=(\frac{a}{\aout})^{3/2}\frac{1}{(1-\eout^2)^{3/2}}\frac{\mper}{[(m+\mper)m]^{1/2}}~.
\end{align}
The flip fraction as a function of $\aout/a$, or equivalently $\epoct$, is shown in the right panel of Figure \ref{random} for 4 values of $\epCDA$. The presented results are from additional ensembles of simulations with isotropically distributed initial orbital orientations. As can be seen, significant  differences between the DA and the CDA calculations occur when 
\begin{equation}
\epCDA\gtrsim\epoct.
\end{equation}
For comparison, the simulation presented in Figure \ref{flip_aper_10} and the lower panel of Figure \ref{jzmper}, has $\epoct=0.021, \epCDA=0.024$ which are comparable, so the magnitude of the correction terms has a non-negligible effect.
For a given set of masses, the coefficient $\epCDA$ depends stronger on $\aout/a$ than $\epoct$ and for sufficiently large $\aout/a$ we have $\epCDA<\epoct$ and the DA approximation converges with the CDA as seen in the left panel of the figure.

In the calculations presented so far in this section the effects of the ``Initial Phase Correction'' (IPC) and ``Fast Oscillation Component'' (FOC) are usually small and have not been included. These effects become more significant when the perturber is more massive. To demonstrate this, the occurrence of flips as a function of orientation for a very massive tertiary, $\mper/m=100$ are shown in Figure \ref{flip_IPC_FOC}. As can be seen, for such massive perturbers, the IPC and FOC corrections are important.

\section{Discussion}

Several previous authors have noted that the effects of short-term oscillations (on the outer orbital time-scale) on the Lidov-Kozai cycles are not captured by the double-averaging (DA) approximation \citep[e.g.][]{kozaibh2,bodewegg,katzdong}. In this paper we demonstrate that the short-term errors can accumulate over time, and the accumulated errors can significantly affect the long-term evolution of the system especially when the mass of the perturber is comparable or larger than that of the inner binary. In particular, the long-term evolution in the Lidov-Kozai cycles due to the octupole effect \citep{ford00,naoz11,katz11,lithwicknaoz11} can be significantly affected by these errors, and as a result, the criteria for achieving orbital ``flips'', where the mutual inclination between the inner and outer orbit crosses $90$ degrees leading to extreme eccentricities, can be considerably
modified from the DA calculations (see Section \S\ref{section:statistics}). 

The leading corrections to the secular equations due to the short-term oscillations in the test particle approximation are derived in \S\ref{sec:ShortTermOscCDA}. This is done by first deriving analytic expressions for the short-term oscillations (Eqs. \eqref{eq:jebar1} and \eqref{eq:djedf_coeff}) and then incorporating them in the outer orbit averaging. The scale of the leading-order correction is set by the small parameter $\epCDA$ defined in Eq. \eqref{eq:epCDA}, which is roughly equal to the amplitudes of the variations of $e,j$ within an outer orbit.
The resulting corrected double-averaging (CDA) equations [Eqs. \eqref{eq:fullCDAterms}- \eqref{eq:fullDA}] are equivalent to adding the correction potential  Eq. \eqref{eq:CDAPotential} to the standard doubly-averaged potential Eq. \eqref{eq:QDA}-\eqref{eq:epoct}. For the limiting case of $\eout=0$ and $\mper\gg m$, the potential reduces to that derived in \citet[][see however issue with  $\omega$ mentioned below Eq. \eqref{eq:CDAPotential}]{Cuk04}. The first-order corrections to the lunar precession are calculated in \S\ref{Appendix:moon} and shown to agree with previous results. The equations are implemented in computer codes which are provided in the supplementary material and described in \S\ref{sec:Implementation}. These can be used as standalone integration codes or added to existing secular codes. 
 
 As shown in \S\ref{section:statistics} and Appendix \S\ref{Appendix:N-body}, the corrected equations capture most of the significant deviation between the N-body integrations and the double-averaging integrations for a broad range of parameters. As can be seen in Figure \ref{flip_aper_5},  at sufficiently low hierarchies, even the CDA equations fail to reproduce the N-body result. By examining a few of such runs, we find that even the single-averaged equations fail for this dynamically violent system with $\epCDA\approx0.07$. 
  
 We emphasize that, while the derived correction is in the test particle limit and thus its applicability is restricted to such systems, the long-term accumulation of the short-term errors found in this work occur under more general conditions. We are in the process of extending the derivations to relax the test particle approximation. Within the test particle approximation, these corrections can be added to other effects such as General Relativistic or tidal precession.
  
 The long-term errors demonstrated in this paper may play an important role in a number of astrophysical settings in which the perturbers have considerable mass. Examples include triple star systems, planets in binary systems, binaries orbiting massive black holes and moons. Previous results studying such systems employing the double-averaging approximation may need to be re-examined.

\acknowledgments

{\bf{Acknowledgments}}
We thank Scott Tremaine and Dong Lai for helpful discussions. 
The work is partially carried out during L.L.'s visit at the Weizmann Institute of Sciences (WIS), and we thank WIS for the supporting the visit. L.L. is partially
supported by the Undergraduate Research Training program of
Peking University. This research was partially supported by by the I-CORE Program (1829/12). B.K. was partially supported by the Beracha Foundation. S.D. is supported by ``the Strategic Priority Research Program-The Emergence of Cosmological Structures'' of the Chinese Academy of Sciences (Grant No. XDB09000000) and Project 11573003 supported by NSFC.



\appendix
\section*{Appendix}

\section{Extended parameter scan for flip citeria}\label{Appendix:N-body}
\begin{figure}[hpt!]
\plotone{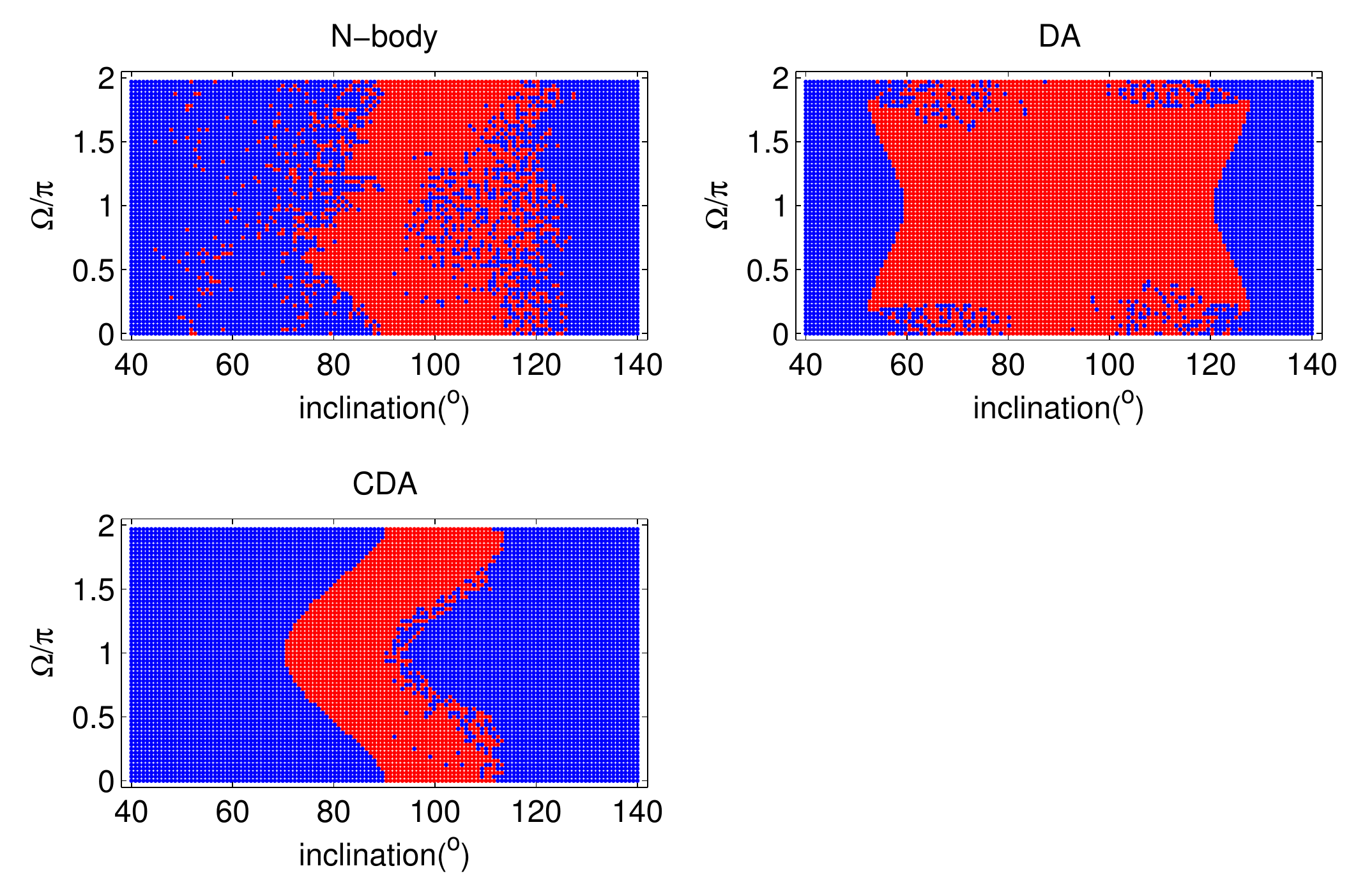}
\caption{Same as Figure \ref{flip_aper_10} except initial $\aout/a=5$.}
\label{flip_aper_5}
\end{figure}

\begin{figure}[hpt!]
\plotone{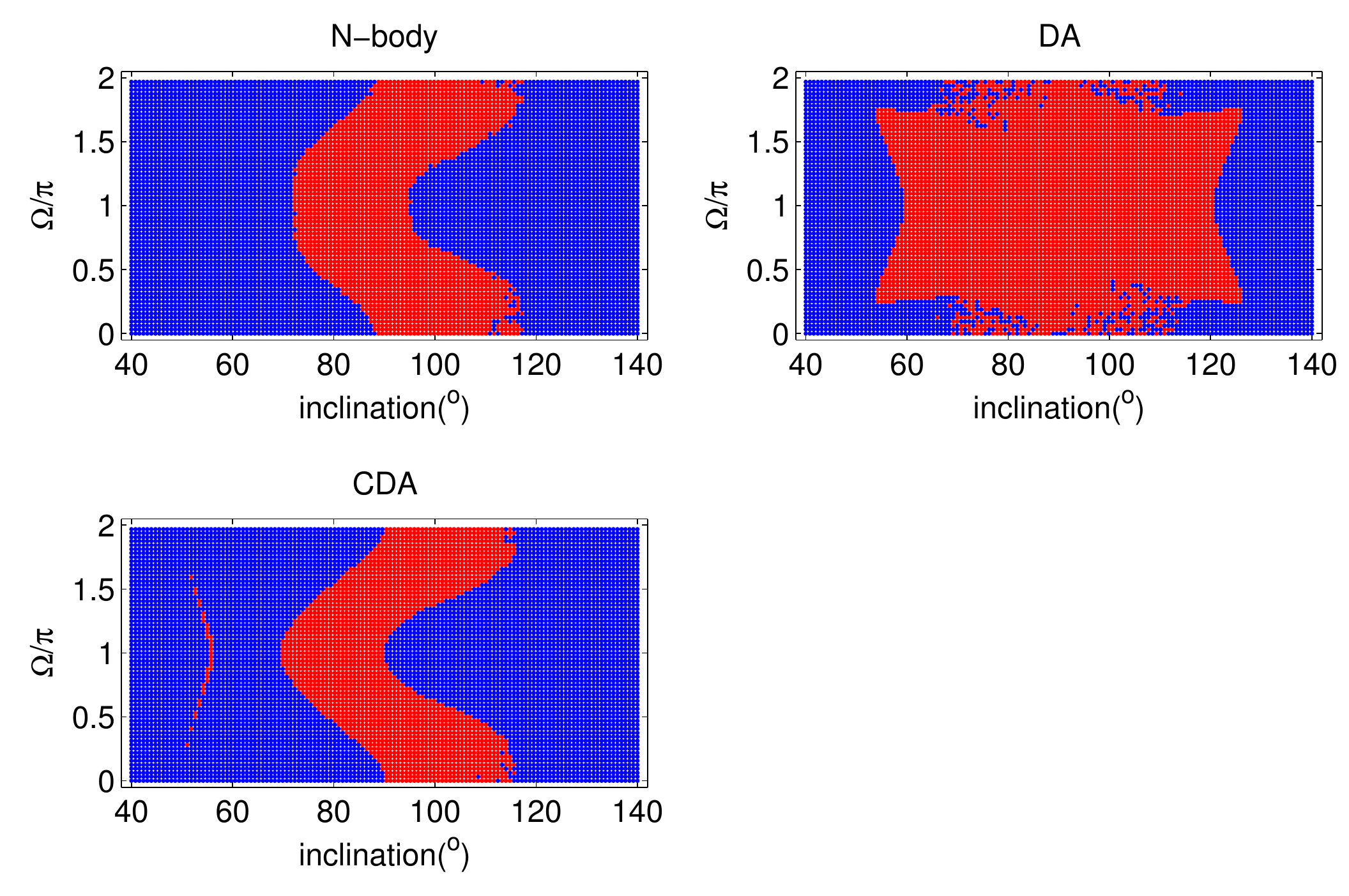}
\caption{Same as Figure \ref{flip_aper_10} except initial $\aout/a=6.5$.}
\label{flip_aper_6d5}
\end{figure}

\begin{figure}[hpt!]
\plotone{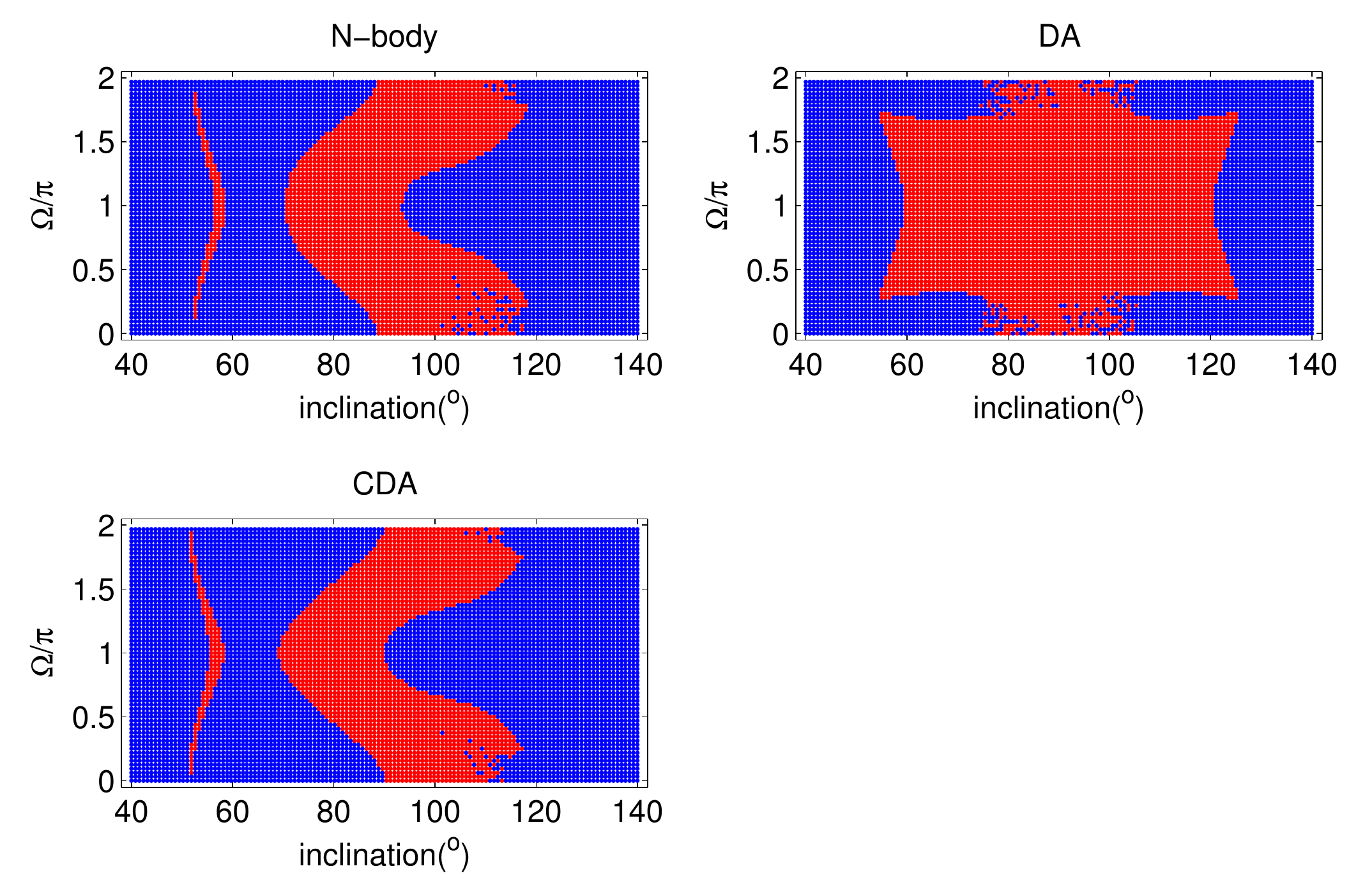}
\caption{Same as Figure \ref{flip_aper_10} except initial $\aout/a=8$.}
\label{flip_aper_8}
\end{figure}

\begin{figure}[hpt!]
\plotone{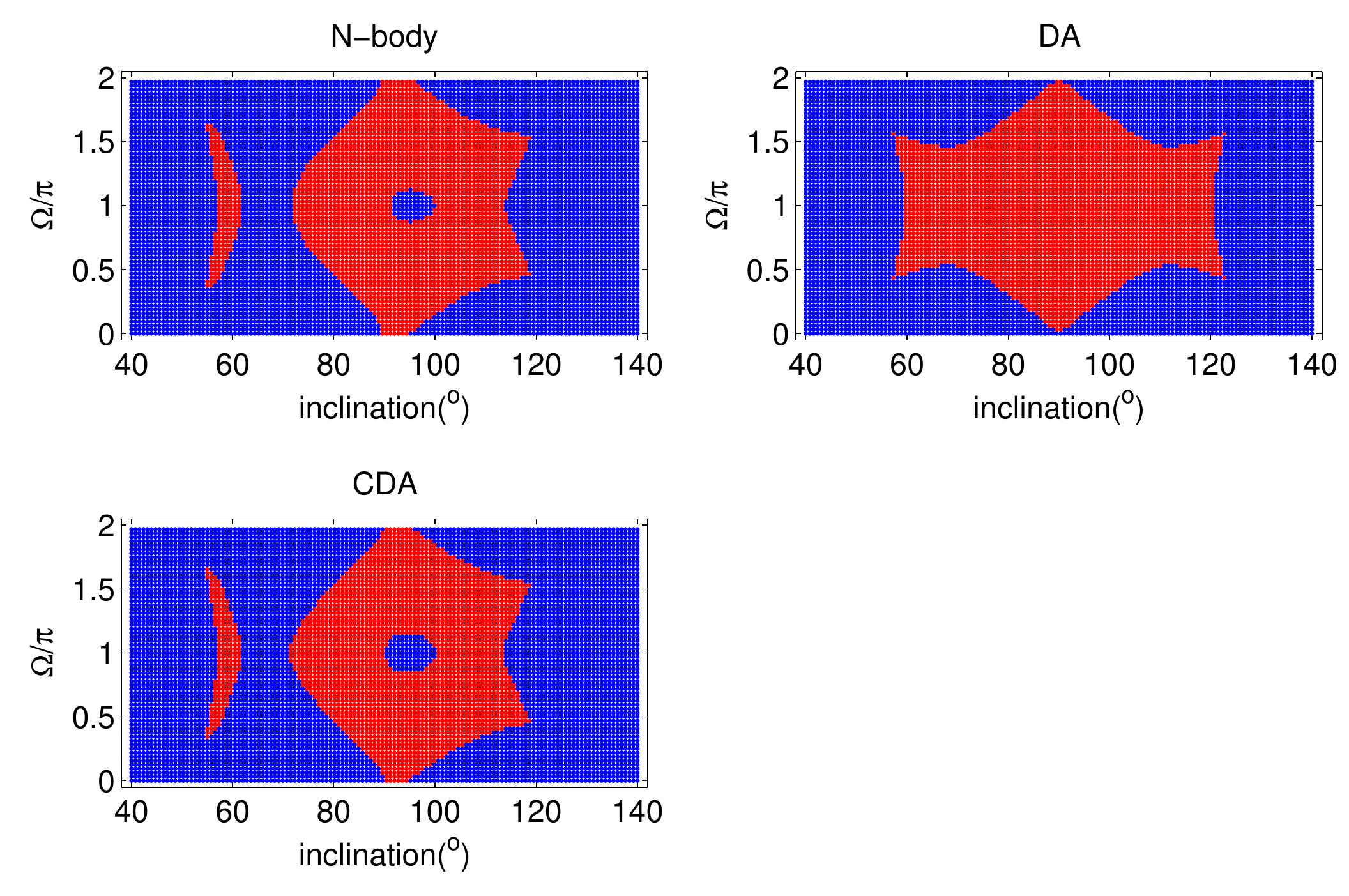}
\caption{Same as Figure \ref{flip_aper_10} except initial $\aout/a=16$.}
\label{flip_aper_16}
\end{figure}

\begin{figure}[hpt!]
\plotone{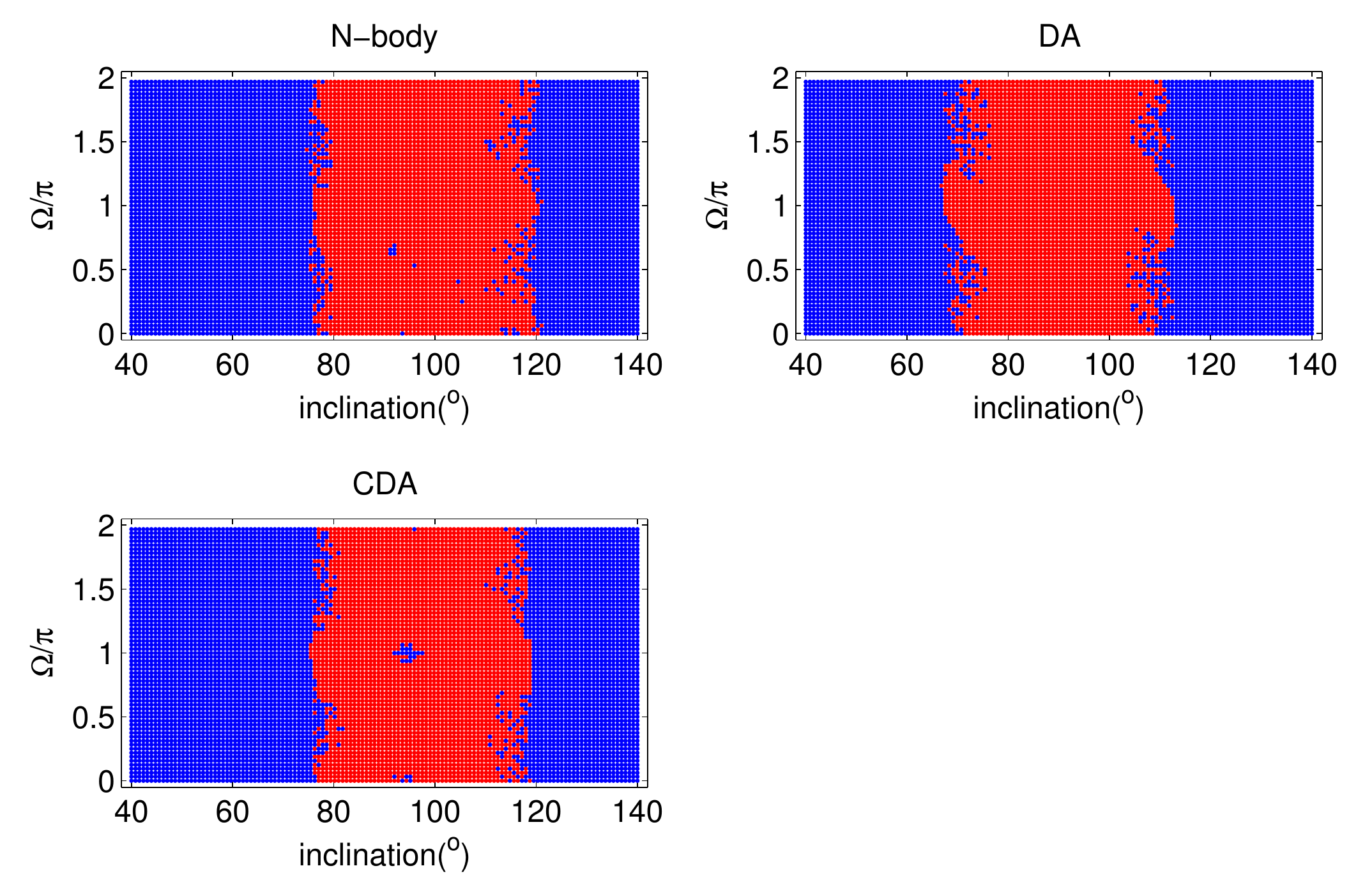}
\caption{Same as Figure \ref{flip_aper_10} except initial $e=0.01$ and $\aout/a=8$.}
\label{flip_aper_8_e_0d01}
\end{figure}

\begin{figure}[hpt!]
\plotone{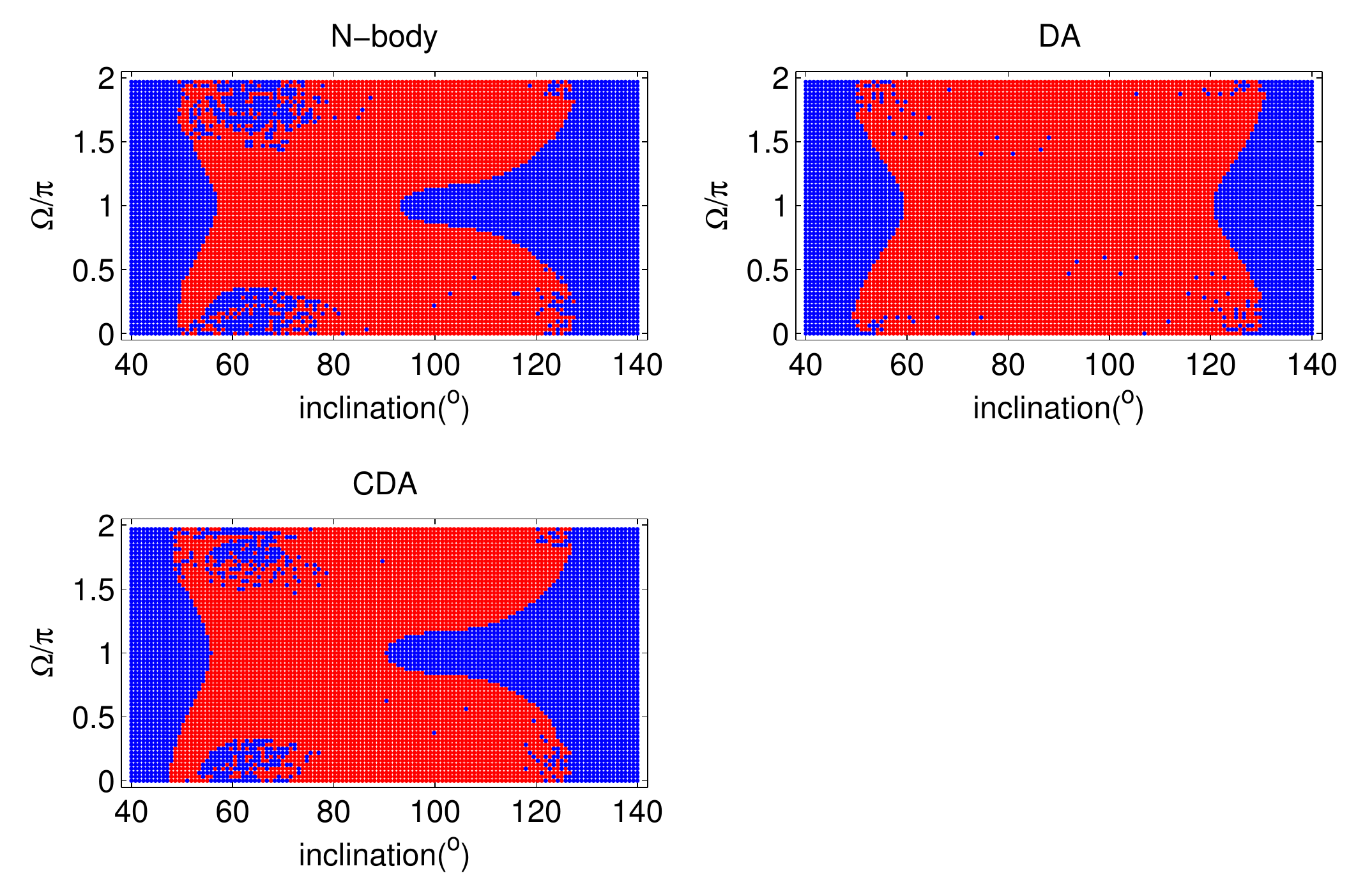}
\caption{Same as Figure \ref{flip_aper_10} except initial $\eout=0.8$ and $\aout/a=30$.}
\label{flip_aper_30_eper_0d8}
\end{figure}

As an expansion of section \S\ref{section:statistics}, the results of additional runs with different initial conditions are provided for  N-body, DA and CDA calculations in Figures \ref{flip_aper_5}-\ref{flip_aper_30_eper_0d8}.
In Figures \ref{flip_aper_5}-\ref{flip_aper_16} the semi-major axis of the outer orbit is varied while all the other parameters are fixed.
Figure  \ref{flip_aper_8_e_0d01}, is the same as Figure \ref{flip_aper_8}, with $\aout/a=8$, but with the inner eccentricity starting at $e=0.01$. Note the significant effect of the small eccentricity.

Figure \ref{flip_aper_30_eper_0d8} shows an example with a high value of the outer eccentricity $\eout=0.8$ (and correspondingly a larger $\aout=30$ to keep the outer orbit's pericenter away from the inner orbit). As can be seen this example, the CDA approximation is also quite reliable demonstrating the fact that the CDA equations are correct for high outer eccentricities.

\newpage
\section{Full analytical expression for Oscillating component}\label{Appendix:fullSAf}
Expressions for  $\mathbfcal{J}_{f,l}^{c},\mathbfcal{J}_{f,l}^{c},\mathbfcal{E}_{f,l}^s,\mathbfcal{E}_{f,l}^c$ which are used in Eq. \eqref{eq:djedfsum} are obtained by expanding Eqs. \eqref{eq:JEflcs}  using Eqs. \eqref{eq:phi_lcs},\eqref{eq:phi_lcs1} and the definition of $\cD_{\bfj},\cD_{\bfe}$ in Eq. \eqref{eq:DjDedef}. The resulting expressions for $l=0,2$ are:
\begin{align}\label{eq:djedf_coeff}
\mathbfcal{J}_{f,0}=\cD_{\bfj}[\phi_0]&=\frac{3}{4}\left(-5 e_y e_z + j_y j_z,~~5 e_x e_z - j_x j_z,~~0\right)~\cr
\mathbfcal{E}_{f,0}=\cD_{\bfe}[\phi_0]&=
\frac{3}{4}\left(-3 e_z j_y - e_y j_z,~~3 e_z j_x + e_x j_z,~~ 2 e_y j_x - 2 e_x j_y\right)~\cr
\mathbfcal{J}_{f,2}^s=\cD_{\bfj}[\phi_2^s]&=
\frac{3}{4}\left(-5 e_x e_z+j_x j_z,~~5 e_y e_z-j_y j_z,~~5 e_x^2-5 e_y^2-j_x^2+j_y^2\right)~,\nonumber\\
\mathbfcal{E}_{f,2}^s=\cD_{\bfe}[\phi_2^s]&=
\frac{3}{4}\left(e_z j_x - 5 e_x j_z,~~-e_z j_y + 5 e_y j_z,~~ 4 e_x j_x - 4 e_y j_y\right)~,\cr
\mathbfcal{J}_{f,2}^c=\cD_{\bfj}[\phi_2^c]&=
\frac{3}{4}\left(5 e_y e_z - j_y j_z,~~5 e_x e_z - j_x j_z,~~-10 e_x e_y + 2 j_x j_y)\right)~,\cr
\mathbfcal{E}_{f,2}^c=\cD_{\bfe}[\phi_2^c]&=
\frac{3}{4}\left(-e_z j_y + 5 e_y j_z,~~-e_z j_x + 5 e_x j_z,~~-4 e_y j_x - 4 e_x j_y\right)~.
\end{align}
The coefficients with $l=1,3$ can be expressed as combinations the coefficients with $l=0,2$ using Eq. \eqref{eq:phi_lcs1}:
\begin{align}\label{eq:phi_lcsA}
&\mathbfcal{J}_{f,1}^c=\eout(\mathbfcal{J}_{f,0}+\mathbfcal{J}_{f,2}^c/2),\cr
&\mathbfcal{E}_{f,1}^c=\eout(\mathbfcal{E}_{f,0}+\mathbfcal{E}_{f,2}^c/2),\cr
&\mathbfcal{J}_{f,3}^c=\eout\mathbfcal{J}_{f,2}^c/2,\cr
&\mathbfcal{E}_{f,3}^c=\eout\mathbfcal{E}_{f,2}^c/2,\cr
&\mathbfcal{J}_{f,1}^s=\mathbfcal{J}_{f,3}^s=\eout\mathbfcal{J}_{f,2}^s/2,\cr
&\mathbfcal{E}_{f,1}^s=\mathbfcal{E}_{f,3}^s=\eout\mathbfcal{E}_{f,2}^s/2.\cr
\end{align}

\section{Full analytic expression for the Corrected doubly-averaged equations}\label{Appendix:fullCDA}

The secular equations of motion, including the double-averaging terms (quadrupole and octupole) and the long-term corrections due to the oscillations discussed in \S\ref{sec:CDA}, can be written as:
\begin{align}\label{eq:fullCDAterms}
\frac{d\sbfj}{d\tau}=&\big(\frac{d\sbfj}{d\tau}\big)^{(DA)}_{Quad}+
\epoct\big(\frac{d\sbfj}{d\tau}\big)^{(DA)}_{Oct}
+\epCDA\big(\frac{d\sbfj}{d\tau}\big)_{\epCDA}
+\epCDA \eout^2\big(\frac{d\sbfj}{d\tau}\big)_{\epCDA \eout^2}~,\nonumber\\
\frac{d\bfe}{d\tau}=&\big(\frac{d\bfe}{d\tau}\big)^{(DA)}_{Quad}+
\epoct\big(\frac{d\bfe}{d\tau}\big)^{(DA)}_{Oct}
+\epCDA\big(\frac{d\bfe}{d\tau}\big)_{\epCDA}
+\epCDA \eout^2\big(\frac{d\bfe}{d\tau}\big)_{\epCDA \eout^2}~,
\end{align}
where $\tau=t/\tsec$ is the secular time [Eq. \eqref{eq:deftau}], $\epoct=[(m_1-m_2)/m](a/\aout)\eout/(1-\eout^2)$ and were we use non-bared symbols $\bfj,\bfe$ instead of the bared symbols $\bar \bfj, \bar \bfe$ for brevity. The last two terms in Eq. \eqref{eq:fullCDAterms} are the new corrections due to short-term (perturber-period) oscillations and are given by

\begin{align}\label{eq:fullCDA}
\big(\frac{d\sbfj}{d\tau}\big)_{\epCDA}=&
\left(\frac{27}{64}(-10e_y e_z j_z+j_y(\frac{1}{3}+8e_x^2+8e_y^2+3e_z^2-j_z^2)),\right. \nonumber \\
&\left.-\frac{27}{64}(-10e_x e_z j_z+j_x(\frac{1}{3}+8e_x^2+8e_y^2+3e_z^2-j_z^2)),0\right)
\nonumber \\
\big(\frac{d\sbfj}{d\tau}\big)_{\epCDA \eout^2}=&
\left(-\frac{9}{64}(10e_y e_z j_z+j_y(-\frac{2}{3}-21e_x^2+9e_y^2-16e_z^2-j_x^2+j_y^2),\right.\nonumber\\
&\frac{9}{64}(30e_x e_z j_z+j_x(\frac{10}{3}-45e_x^2-15e_y^2-5j_x^2-3j_y^2)), \nonumber \\
&\left.-\frac{9}{16}(5e_y e_z j_x+5e_x e_z j_y+5e_x e_y j_z+j_x j_y j_z)\right)
\nonumber \\
\big(\frac{d\bfe}{d\tau}\big)_{\epCDA}=&
\left(\frac{27}{64}(6e_z j_y j_z+e_y(\frac{1}{3}+8e_x^2+8e_y^2+3e_z^2-17j_z^2)),\right. \nonumber \\
&-\frac{27}{64}(6e_z j_x j_z+e_x(\frac{1}{3}+8e_x^2+8e_y^2+3e_z^2-17j_z^2)), \nonumber \\
&\left.\frac{27}{4} (e_y j_x-e_x j_y) j_z\right)
\nonumber \\
\big(\frac{d\bfe}{d\tau}\big)_{\epCDA \eout^2}=&
\left(\frac{9}{64}(14e_z j_y j_z+e_y(\frac{35}{3}+10e_x^2+5e_z^2-10j_x^2-32j_y^2-35j_z^2)), \right. \nonumber \\
&-\frac{9}{64}(10e_z j_x j_z+e_x(\frac{65}{3}-10e_y^2-25e_z^2-22j_y^2-65j_z^2)), \nonumber \\
&\left.-\frac{9}{16} (5 e_x e_y e_z+5 e_z j_x j_y-5 e_y j_x j_z+11 e_x j_y j_z)\right).
\end{align}
The first two terms in each of Eqs. \eqref{eq:fullCDAterms} are the (previously known) double-averaging quadrupole and octupole contributions and are given by \citep[e.g.][]{katzdong}
\begin{align}\label{eq:fullDA}
\big(\frac{d\sbfj}{d\tau}\big)^{(DA)}_{Quad}=&
\left(\frac{3}{4} (-5 e_y e_z+j_y j_z),\frac{3}{4} (5 e_x e_z-j_x j_z),0\right),
\nonumber \\
\big(\frac{d\sbfj}{d\tau}\big)^{(DA)}_{Oct}=&
\left(-\frac{75}{32} (-7 e_x e_y e_z+e_z j_x j_y+e_y j_x j_z+e_x j_y j_z), \right. \nonumber \\
&\frac{15}{64} (20 e_x j_x j_z+e_z (1-78 e_x^2-8 e_y^2+27 e_z^2+10 j_x^2-15 j_z^2)), \nonumber \\
&\left.\frac{15}{64} (10 e_z j_y j_z+e_y (-1+8 e_x^2+8 e_y^2-27 e_z^2+5 j_z^2))\right),
\nonumber \\
\big(\frac{d\bfe}{d\tau}\big)^{(DA)}_{Quad}=&
\left(-\frac{3}{4} (3 e_z j_y+e_y j_z),\frac{3}{4} (3 e_z j_x+e_x j_z),\frac{3}{2} (e_y j_x-e_x j_y)\right),
\nonumber \\
\big(\frac{d\bfe}{d\tau}\big)^{(DA)}_{Oct}=&
\left(\frac{15}{32} (-5 e_y e_z j_x+27 e_x e_z j_y+3 e_x e_y j_z-5 j_x j_y j_z), \right.\nonumber \\
&-\frac{15}{64} (44 e_x e_z j_x+j_z (-1+14 e_x^2+8 e_y^2-17 e_z^2-10 j_x^2+5 j_z^2)), \nonumber \\
&\left.\frac{15}{64} (26 e_y e_z j_z+j_y (-1+24 e_x^2+24 e_y^2-27 e_z^2+5 j_z^2))\right).
\nonumber \\
\end{align}

\section{Application to the precession of the moon's orbit}\label{Appendix:moon}
In this section the CDA approximation is applied to the precession of the moon's orbit to verify that it reproduces the correct (known) precession rate to leading order.

In the moon-earth-sun system, the orbit of the moon is perturbed by the sun. The moon can be treated as a test particle. For the Earth-moon-sun system we have $\tsec=2.1$ years, $\epCDA=0.075$ and $\epoct=4.1\times10^{-5}$. The octuple and the $\epCDA\eout^2$ correction terms are negligible.

The moon's orbit has small eccentricity ($e\approx0.05\ll1$) and small inclination with respect to ecliptic plane ($\approx5^\circ$), so to the first order in $e_x,e_y,e_z,j_x,j_y$ and applying $j_z\approx1$ we obtain from Eqs. \eqref{eq:fullCDAterms},\eqref{eq:fullDA}
\begin{align}\label{eq:Moonje}
\frac{dj_x}{d\tau}&=(\frac{3}{4}-\frac{9}{32}\epCDA) j_y~, \nonumber\\
\frac{dj_y}{d\tau}&=-(\frac{3}{4}-\frac{9}{32}\epCDA) j_x~. \\
\frac{de_x}{d\tau}&=-(\frac{3}{4}+\frac{225}{32}\epCDA)e_y~, \nonumber\\
\frac{de_y}{d\tau}&=(\frac{3}{4}+\frac{225}{32}\epCDA)e_x~,
\end{align}
which imply a nodal precession rate of
\begin{equation}
n_{\rm nodal}=\frac{1}{\tsec}\left(\frac{3}{4}-\frac{9}{32}\epCDA\right)
\end{equation}
and apsidal precession rate of 
\begin{equation}
n_{\rm apsidal}=\frac{1}{\tsec}\left(\frac{3}{4}+\frac{225}{32}\epCDA\right)
\end{equation}
in agreement with the known first order corrections at low inclination and eccentricity (see e.g. \cite{Cuk04} and references therein).

If we use the doubly-averaged equations, setting $\epCDA=0$ we find that the periods of the nodal precession and apsidal precession have the same period of $2\pi \tsec/\frac{3}{4}\approx17.7$ years. The observed nodal precession period is 18.6 years - 5\% accuracy; but the observed apsidal precession period is 8.9 years - factor of 2 off.  Leonhard Euler, Alexis Clairaut and Jean d'Alembert obtained the same puzzling result and were seriously considering the possibility that the $1/r^2$ gravitational law is wrong.

The first order correction in  $\epCDA$ leads to a small correction to the nodal precession: 17.7 $\rightarrow$ 18.2 years; but a large correction to the apsidal precession: 17.7 $\rightarrow$ 10.4 years, bridging most of the gap to the actual precession rate (leaving 17\% error due to higher order terms). The reason that there is a significant correction even though the expansion parameter $\epCDA$ is small, is due to the large pre-factor to the relative contribution of the first order correction, $\frac{225}{32}\frac{4/3}\sim 10$.


\end{document}